%% file: main.tex
\documentclass{new_tlp}

\input{macros}

\usepackage{mathptmx}
\usepackage{enumitem}

\usepackage{soul}

  \title[Practical Reasoning in DatalogMTL]
        {Practical Reasoning in DatalogMTL
        }

  \author[D. Wang et al.]
{DINGMIN WANG \\ 
	Department of Computer Science, University of Oxford, UK\\
	\email{dingmin.wang@cs.ox.ac.uk}\\
	\and   PRZEMYSŁAW A. WAŁĘGA \\ 
	Department of Computer Science, University of Oxford, UK\\
	\email{przemyslaw.walega@cs.ox.ac.uk} \\ 
	\and   PAN HU \\ 
	School of Electronic Information and Electrical Engineering, Shanghai Jiao Tong University, China\\
	\email{pan.hu@sjtu.edu.cn} \\ 
	\and  BERNARDO CUENCA GRAU \\ 
	Department of Computer Science, University of Oxford, UK\\
	\email{bernardo.cuenca.grau@cs.ox.ac.uk}
}

\jdate{March 2003}
\pubyear{2003}
\pagerange{\pageref{firstpage}--\pageref{lastpage}}
\doi{S1471068401001193}

\newtheorem{theorem}{Theorem}[section]
\newtheorem{lemma}[theorem]{Lemma}

\newtheorem{example}[theorem]{Example}
\newtheorem{definition}[theorem]{Definition}

\begin{document}

\label{firstpage}
 
\maketitle

\begin{abstract}
\MTL{} is an extension of Datalog with metric temporal operators that has found an increasing number of applications
in recent years. Reasoning in \MTL{} is, however, of high computational complexity, 
which makes reasoning in modern data-intensive applications challenging.
In this paper we present a practical reasoning algorithm for the full \MTL{} language, which we have implemented in a
system called MeTeoR. Our approach
effectively combines an optimised (but generally non-terminating) materialisation (a.k.a.\ forward chaining)
procedure, which provides scalable behaviour, with an automata-based component that guarantees termination and completeness.
To ensure favourable scalability of the materialisation component, 
we propose a novel semina\"ive materialisation procedure for \MTL{} enjoying the non-repetition property, which
ensures that  each specific rule application will be considered at most once throughout the entire execution of the algorithm. Moreover, our materialisation procedure is enhanced
with additional optimisations which 
further reduce the number of redundant computations performed 
during materialisation by disregarding 
rules  as soon as it is certain
that they cannot derive new facts in subsequent 
materialisation steps.
Our extensive evaluation supports the practicality of our approach.
\end{abstract}


\input{sections/introduction}

\input{sections/related}

\input{sections/preliminary}

\input{sections/techniques}

\input{sections/practical}

\input{sections/experiments}

\input{sections/conclusions}
		
\section*{Acknowledgments}

Our research  was funded  by the following
 EPSRC projects: OASIS (EP/S032347/1),
and UK FIRES (EP/S019111/1),
as well as by SIRIUS Centre for Scalable Data Access, 
Samsung Research UK, and NSFC grant No. 62206169.
For the purpose of Open Access, the authors have applied a CC BY public copyright licence to any Author Accepted Manuscript (AAM) version arising from this submission.

\bibliographystyle{acmtrans}
\bibliography{reference}

\newpage
\appendix

\input{sections/appendix}

\label{lastpage}
\end{document}

%% file: macros.tex
\usepackage{amssymb}
\usepackage{mathtools}
\usepackage[T1]{fontenc}
\usepackage[utf8]{inputenc}

\usepackage{url}

\usepackage{cleveref}
\usepackage{stmaryrd}
\usepackage{graphicx}
\usepackage[ruled,vlined,linesnumbered]{algorithm2e}
\Crefname{algocf}{Algorithm}{Procedure}
\SetAlgorithmName{Procedure}{}{}

\usepackage{paralist}

\usepackage{tikz}
\usepackage{pgfplots}
\usetikzlibrary{arrows,backgrounds,automata,topaths,patterns,decorations.pathreplacing,decorations.pathmorphing}
\usetikzlibrary{arrows, chains, positioning, quotes, shapes.geometric}
\usetikzlibrary{calc}
\usetikzlibrary{fit}
\usetikzlibrary{arrows}
\usetikzlibrary{positioning,automata}

\newcommand{\MTL}{\ensuremath{\textup{DatalogMTL}}}
\newcommand{\matA}{M}

\newcommand{\sbf}{\mathbf{s}}
\newcommand{\cbf}{\mathbf{c}}
\newcommand{\xbf}{\mathbf{x}}

\newcommand{\PS}{\ensuremath{\textsc{\textup{PSpace}}}}
\newcommand{\EXPS}{\ensuremath{\textsc{\textup{ExpSpace}}}}

\DeclareFontFamily{U}{MnSymbolC}{}
\DeclareSymbolFont{MnSyC}{U}{MnSymbolC}{m}{n}

\DeclareMathSymbol{\square}{\mathbin}{MnSyC}{106}
\DeclareMathSymbol{\meddiamond}{\mathbin}{MnSyC}{110}
\DeclareMathSymbol{\boxminus}{\mathbin}{MnSyC}{112}
\DeclareMathSymbol{\boxplus}{\mathbin}{MnSyC}{116}
\DeclareMathSymbol{\diamondminus}{\mathbin}{MnSyC}{120}
\DeclareMathSymbol{\diamondplus}{\mathbin}{MnSyC}{124}

\DeclareFontShape{U}{MnSymbolC}{m}{n}{
	<-5>  MnSymbolC4
	<5-6>  MnSymbolC5
	<6-7>  MnSymbolC6
	<7-8>  MnSymbolC7
	<8-9>  MnSymbolC8
	<9-10> MnSymbolC9
	<10-12> MnSymbolC10
	<12->   MnSymbolC12}{}

\newcommand{\Si}{\mathcal{S}}
\newcommand{\Ui}{\mathcal{U}}

\newcommand{\I}{\mathfrak{I}}
\newcommand{\D}{\mathcal{D}}
\newcommand{\Prog}{\Pi}
\newcommand{\ins}[2]{\ensuremath{\mathsf{inst}_{#1}[#2]}}
\newcommand{\der}[2]{\ensuremath{#1[#2]}}
\newcommand{\semi}[3]{\ensuremath{#1[ #2 \threedotcolon #3  ]}}

\newcommand{\can}[2]{\mathfrak{C}_{#1,#2}}
\newcommand{\ground}[2]{\mathsf{ground}(#1,#2)}

\newcommand{\coal}[1]{\mathsf{coal}(#1)}

\newcommand{\true}{\mathsf{true}}
\newcommand{\false}{\mathsf{false}}

\newcommand{\N}{\mathbb{N}}
\newcommand{\Q}{\mathbb{Q}}
\newcommand{\Z}{\mathbb{Z}}

\newcommand{\inc}{\mathsf{inconst}}

\newcommand{\threedotcolon}{\,\substack{\cdot\\[-0.12cm]\cdot\\[-0.12cm]\cdot\\[0.05cm]}\,}

\newcommand{\subtag}[1]{\tag{#1}}

\newcommand{\ltl}{\textbf{LTL}}

\definecolor{LightGray}{RGB}{150,150,150}
\definecolor{Green1}{RGB}{171,255,177}
\definecolor{Green2}{RGB}{65,243,77}
\definecolor{Green3}{RGB}{120,165,101}
\definecolor{Brown}{RGB}{243,190,130}
\definecolor{Blue1}{RGB}{93,114,164}
\definecolor{Blue2}{RGB}{213,223,245}
\definecolor{Blue3}{RGB}{42,69,176}
\definecolor{Blue4}{RGB}{240,246,255}
\definecolor{Blue5}{RGB}{50,100,200}
\definecolor{Red1}{RGB}{255,89,100}
\definecolor{Light}{RGB}{102,205,170}

%% file: sections/introduction.tex
\section{Introduction}
\MTL{} \cite{brandt2018querying} is an extension
of the core rule-based  language Datalog~\cite{ceri1989you} with operators from
metric temporal logic (MTL)~\cite{koymans1990specifying} interpreted
over the rational timeline. 
For example, the following \MTL{} rule states that travellers can enter the US if 
they had a negative COVID-19 test sometime  in the last $2$ days ($\diamondminus_{[0,2]}$) and have held fully vaccinated status continuously 
throughout the last $15$ days ($\boxminus_{[0,15]}$)
\begin{align*}
\mathit{Authorised(x)}\gets \diamondminus_{[0,2]}\mathit{NegativeLFT(x)} \land \boxminus_{[0,15]}\mathit{FullyVaccinated(x)}.
\end{align*}
\MTL{} is a  powerful temporal knowledge representation language,
which has recently found applications
in ontology-based query answering~\cite{brandt2018querying,kikot2018data,guzel2018ontop,koopmann2019ontology} and stream
reasoning~\cite{walega2019reasoning}, amongst others \cite{DBLP:conf/edbt/NisslS22,mori2022neural}. 
Reasoning in \MTL{} is, however, of
high complexity, namely \EXPS-complete \cite{brandt2018querying} and
\PS-complete with respect to data size  \cite{walega2019datalogmtl},
which makes reasoning in data-intensive applications challenging.
Thus, theoretical research has focused on 
establishing a suitable trade-off between expressive power and complexity of reasoning,
by 
identifying lower complexity fragments
of \MTL{} \cite{DBLP:conf/ijcai/WalegaGKK20,walkega2023finite} and studying 
alternative semantics with favourable computational behaviour
\cite{DBLP:conf/kr/WalegaGKK20,ryzhikov2019data}.

The design and implementation of practical reasoning algorithms for \MTL{}  remains, however,
a largely unexplored area---something
that has  so far prevented its widespread adoption in applications.
\citeN{brandt2018querying} implemented a prototype reasoner based on 
query rewriting that is applicable only to 
non-recursive \MTL{} programs; in turn, the temporal extension of
the Vadalog system \cite{DBLP:journals/pvldb/BellomariniSG18} described by \citeN{tempVada} implements the
full \MTL{} language, but without termination guarantees.
The lack of available reasoners for \MTL{} is in stark contrast with  plain
Datalog, and closely related Answer Set Programming,
for which a plethora of (both academic and commercial) systems have been developed and successfully deployed
in practice
\cite{motik2014parallel,DBLP:conf/semweb/CarralDGJKU19,DBLP:journals/pvldb/BellomariniSG18,DBLP:journals/corr/GebserKKS14,DBLP:conf/kr/EiterLMPS98,DBLP:journals/ai/GebserKS12}.

In this paper, we present the first practical reasoning algorithm for the full \MTL{} language interpreted under the standard semantics (so called continuous semantics over the rational timeline), which is
sound, complete, and terminating. 
Our algorithm
combines materialisation (a.k.a.\ forward chaining) and automata-based reasoning.
On the one hand, materialisation is the reasoning paradigm of choice in numerous  
Datalog systems \cite{DBLP:conf/rweb/BryEEFGLLPW07,DBLP:conf/rweb/BryEEFGLLPW07,motik2014parallel,DBLP:conf/semweb/CarralDGJKU19,DBLP:journals/pvldb/BellomariniSG18}; 
facts logically entailed by an input program and dataset are derived in successive rounds of rule applications (also called \emph{materialisation steps})
until  a fixpoint is reached or a fact of interest (or a contradiction) is derived; both this process and its output are often referred to as \emph{materialisation}. 
A direct implementation of materialisation-based
reasoning in \MTL{} is, however, problematic  since
forward chaining may require infinitely many
rounds of rule applications \cite{walega2021finitely,walkega2023finite}. 
An alternative to materialisation-based reasoning ensuring completeness and termination
relies on 
 constructing
B\"{u}chi automata  and checking non-emptiness of their languages 
\citeN{walega2019datalogmtl}. This procedure, however, was mainly proposed
for obtaining tight  complexity bounds,  and not with efficient
implementation in mind;
in particular, the constructed automata are of exponential size, 
which makes direct  implementations
impractical.
Our approach deals with these  difficulties by 
providing optimised materialisation-based algorithms together with an effective way of combining the scalability of
materialisation-based reasoning and the completeness guaranteed by automata-based procedures, 
thus providing `the best of both worlds'. 

After discussing related work in Section~\ref{sec:related} and preliminary definitions for \MTL{} in Section~\ref{sec:Preliminaries}, we will present the following  contributions of our work.
\begin{compactenum}
\item 
In Section \ref{sec:background}
 we recapitulate different 
techniques available for reasoning in \MTL{}, which we will take as a starting point for the
development of our approach. On the one hand, we present a variant of
the (non-terminating) na\"ive materialisation procedure \cite{walkega2023finite}, and prove its 
soundness and completeness; on the other hand, we describe existing algorithms
based on exponential reductions to reasoning in linear temporal logic (LTL) and to emptiness checking of B\"uchi
automata.

\item 
In Section \ref{subsec:seminaive} we present a \emph{semina\"ive} materialisation procedure for \MTL{}.
In contrast to the na\"ive procedure provided in Section \ref{sec:naive} and analogously to the classical semina\"ive algorithm for plain Datalog~\cite{abiteboul1995foundations,DBLP:journals/ai/MotikNPH19},  our procedure aims at minimising 
redundant computation by keeping track of newly derived facts in each materialisation step 
and ensuring that rule applications in the following materialisation step involve at least one of these newly derived
facts. 
In this way, each rule instance is considered at most once, and so our procedure is said to enjoy the \emph{non-repetition} property. 

\item 
In Section \ref{subsec:oseminaive} we present an optimised variant of our semina\"ive procedure which 
further reduces the number of redundant computations performed 
during materialisation by disregarding 
rules during  the execution of the procedure as soon as we can be certain 
that their application will never derive new facts in subsequent 
materialisation steps.

\item 
In Section \ref{sec:combined} we propose  a practical  reasoning algorithm 
combining optimised semina\"ive materialisation and 
our realisation of the
automata-based reasoning approach of \citeN{walega2019datalogmtl}.
Our algorithm is designed  to 
delegate  the bulk of the computation to the scalable materialisation
component and resort to automata-based techniques only as needed to
ensure termination and completeness.

\item 
We have implemented our approach  in the MeTeoR (Metric Temporal Reasoner) system, which we have made publicly available.\footnote{See \url{https://github.com/wdimmy/DatalogMTL_Practical_Reasoning} for the version of MeTeoR described and evaluated in this paper.}
In Section~\ref{sec:experiments} we describe our implementation and  present its evaluation
on a temporal extension of the Lehigh University Benchmark (LUBM), the Weather Benchmark used by \citeN{brandt2018querying}, and 
the iTemporal benchmark generator developed by   \citeN{iTemporal}. 
In Section~\ref{sec:exper} we describe the results of our evaluation and draw the following conclusions.
\begin{compactenum} 
\item[--]
Although completeness and termination in our approach  can be ensured by relying on either automata construction or on
a reduction to
LTL satisfiability followed by the application of a LTL reasoner, our experiments show that  
automata construction yields superior performance and can be used to solve non-trivial problems; this justifies our choice of automata over LTL in MeTeoR.


\item[--] 
Our proposed semina\"ive materialisation strategy offers 
significant performance and scalability gains over na\"ive materialisation, 
and it can be successfully applied to non-trivial programs and datasets with tens of millions of temporal facts. 
 
\item[--] 
The performance of MeTeoR is superior to that of the 
query rewriting approach proposed by \citeN{brandt2018querying} on non-recursive programs.

\item[--] 
The vast majority of queries for programs generated  by  iTemporal  can be answered using materialisation only.
This supports the practicality of our approach since the bult of the workload is delegated to the scalable materialisation component and
the use of automata-based reasoning
is only required in exceptional cases.

\end{compactenum}
\end{compactenum}
This paper expands upon our previous conference publications, specifically at AAAI \cite{wang2022meteor} and RuleML+RR \cite{seminaiveRR}. It offers a more extensive range of experiments, investigates the efficacy of methods relying on reduction to LTL satisfiability, and includes comprehensive proofs of our theoretical results in  a dedicated technical appendix.

%% file: sections/related.tex
\section{Related Work}\label{sec:related}

\MTL{}  was first introduced by \citeN{brandt2017ontology}
and it has  found applications in areas as diverse as temporal stream reasoning \cite{walega2019reasoning},
temporal ontology-based data access~\cite{brandt2017ontology}, 
specification and
verification of banking agreements~\cite{DBLP:conf/edbt/NisslS22},
fact-checking economic claims \cite{mori2022neural},
 and the description of human movements~\cite{raheb2017balonse}.
The complexity of
standard reasoning tasks in both the full language and
its fragments 
has been
investigated  in depth \cite{brandt2018querying,walega2019datalogmtl,DBLP:conf/ijcai/WalegaGKK20,bellomarini2021query}.
Alternative semantics to the standard continuous semantics
over the rational timeline have also been considered; these include
the semantics based  
on the integer timeline studied by \citeN{DBLP:conf/kr/WalegaGKK20}
and the event-based semantics by  \citeN{ryzhikov2019data}. 
\MTL{} has also been recently extended with negation-as-failure under the stable model semantics, first for stratified
programs \cite{DBLP:conf/aaai/CucalaWGK21} and
subsequently for the general case 
\cite{walkega2021datalogmtl}.

There have been numerous alternative proposals for extending Datalog
with temporal constructs.
For example, Datalog$_{1S}$ \cite{chomicki1988temporal} is a prominent 
early extension, in which predicates
are allowed to contain terms of an additional temporal sort and a single
successor function symbol over this sort.
A number of recent temporal extensions of Datalog feature operators from LTL \cite{artale2015first} as well as from the Halpern-Shoham logic of intervals \cite{10.5555/3060621.3060782}.
Extensions of \MTL{} with non-monotonic negation are closely related to temporal extensions of answer set programming (ASP)~\cite{aguado2021}.
Particularly relevant is a recently introduced extension of ASP with MTL operators~\cite{cabalar_dieguez_schaub_schuhmann_2020} as well as the
LARS framework~\cite{beck2018lars}, which combines
ASP and MTL operators for reasoning  over data streams.  
It is worth observing, however,  that 
all these temporal extensions of ASP are interpreted over the integer timeline.
Finally, operators from MTL have also been exploited in
temporal extensions of description logics 
\cite{gutierrez2016metric,DBLP:conf/frocos/BaaderBKOT17,thost2018metric,artale1998temporal}.

The first system 
with \MTL{} support was an extension of the Ontop platform~\cite{guzel2018ontop} restricted to non-recursive programs
and  based on  reasoning via rewriting into SQL.
Very recently, the Vadalog reasoning system \cite{DBLP:journals/pvldb/BellomariniSG18} has also
been extended with \MTL{} support \cite{tempVada}; the algorithm
implemented by this extension is, however, non-terminating in general.
Fragments of \MTL{}  for which  materialisation-based reasoning 
is guaranteed
to terminate have been identified and studied by \citeN{walkega2023finite};  these
fragments, however, impose 
restrictions  which effectively disallow
programs expressing `recursion through time'. 
Finally, there is a plethora of reasoners available for solving
satisfiability checking problems in LTL, with techniques involving
reduction to model checking, tableau systems,  and 
automata-based techniques. 
Prominent recent examples
of highly-optimised LTL reasoners include  nuXmv \cite{DBLP:conf/cav/CavadaCDGMMMRT14} and BLACK \cite{DBLP:conf/tableaux/GeattiGM19}, where
the latter was our LTL reasoner of choice in the experiments reported in
Section~\ref{sec:experiments}.

%% file: sections/preliminary.tex
\section{Preliminaries}\label{sec:Preliminaries}
 
In this section, we recapitulate the syntax, semantics, and key reasoning problems in \MTL{}.

\subsection{Syntax}\label{sec:syntax}

We consider a \emph{timeline} consisting of ordered rational numbers, denoted by $\mathbb{Q}$, which we also call \emph{time points}.
A \emph{(rational) interval} $\varrho$ is a set of rational numbers that is either empty or such that
\begin{itemize}
\item[--] for all $t_1,t_2, t_3 \in \mathbb{Q}$ satisfying 
$t_1<t_2<t_3$ and $t_1,t_3 \in \varrho$, it must be the case that $t_2 \in\varrho$, and
\item[--] the greatest lower bound $\varrho^-$ and the least upper bound $\varrho^+$ of $\varrho$  belong to $\mathbb{Q} \cup \{-\infty ,\infty\}$.\footnote{
By this restriction, for example, the set of all rational numbers between $0$ and $\pi$ is not an interval, as it has no least upper bound in $\Q$.
Also note that in
contrast to \citeN{brandt2018querying}, we do not require finite  endpoints of intervals to
be dyadic rational numbers.}
\end{itemize}
The bounds $\varrho^-$ and $\varrho^+$ are called the \emph{left} and the \emph{right endpoints} of $\varrho$, respectively.
An interval is  \emph{punctual} if it contains exactly one time point. 
Intervals
$\varrho_1$ and $\varrho_2$ are \emph{union-compatible} if $\varrho = \varrho_1 \cup \varrho_2$ is also an interval.
We represent a non-empty interval $\varrho$ using the standard
notation $\langle \varrho^-,\varrho^+ \rangle$, where  
the  \emph{left bracket}
`$\langle$'  is either  `$[$' or `$($', and the \emph{right bracket} $\rangle$ is
either `$]$' or `$)$'.
We assume that each rational endpoint is
given as a (not necessary reduced) fraction with an integer numerator and a~positive integer denominator, both encoded in binary. 
As usual, the brackets `$[$'  and `$]$' indicate that the
corresponding endpoints are included in the interval, whereas `$($' and 
`$)$' indicate that they are not included. 
Observe that every interval has multiple representations due to possibly non-reduced fractions.
Each representation, however, uniquely determines
an interval and so, if it is clear from the context, we will abuse
notation and identify an interval representation with the unique
interval it represents.

We consider a \emph{signature} consisting
of pairwise disjoint countable sets of constants, variables,  and predicates with  
 non-negative integer arities. As usual, a term is either a constant or a variable.
A \emph{relational atom} is an expression of the form $P(\sbf)$,
with $P$ a predicate and $\sbf$ a tuple of terms whose length matches the arity of $P$.
A \emph{metric atom} is an expression
given by the following grammar, where $P(\sbf)$ is a relational atom, $\top$ and $\bot$
are logical truth and falsehood respectively, and
$\diamondminus$ (`sometime in the past'), $\diamondplus$ (`sometime in the future'), $\boxminus$ ('always in the past'), $\boxplus$ ('always in the future'), $\Si$ ('since'), and $\Ui$ ('until') are MTL
operators that can be indexed with any intervals $\varrho$ containing only non-negative rationals:
\begin{equation*}
M  \Coloneqq
\top  \mid  \bot  \mid
P(\sbf)  \mid
\diamondminus_\varrho M   \mid
\diamondplus_\varrho M   \mid   
\boxminus_\varrho M   \mid
\boxplus_\varrho M  \mid
M \Si_\varrho M   \mid
M \Ui_\varrho M.
\end{equation*}
We  refer to $\diamondminus$, $\boxminus$, and $\Si$ as \emph{past operators} and we refer to  $\diamondplus$, $\boxplus$, and $\Ui$ as \emph{future operators}. 
A \emph{rule} is an expression of the form 
\begin{equation}
M'  \leftarrow \matA_1 \land \dots \land \matA_n, \quad  \text{ for } n \geq 1, \label{eq:rule}
\end{equation}
with each $\matA_i$ a metric atom, and $\matA'$  generated by the following grammar:
%
\begin{equation*}
	\matA'   \Coloneqq
	 \bot \mid P(\sbf)  \mid
	\boxminus_\varrho M'   \mid
	\boxplus_\varrho M'. 
\end{equation*}
The conjunction $\matA_1 \land
\dots \land \matA_n$ in Expression~\eqref{eq:rule} is the rule's \emph{body} and $\matA'$ is the rule's \emph{head}. 
A rule mentioning $\bot$ in the head is referred to as a \emph{$\bot$-rule}.
A rule is \emph{forward-propagating} (respectively \emph{backwards-propagating}) if it  does not mention $\top$ or $\bot$,  mentions only past (respectively, future) 
operators in the body, and only future (respectively, past) operators  in the head.
A rule is \emph{safe}
if each variable
in its head also occurs in the body, and this occurrence is not in a
left operand of $\Si$ or $\Ui$; for instance, 
a rule such as $P(x) \gets Q(x) \Si_{[0,0]} \top$ is not safe; in fact, it  can be
equivalently rewritten as $P(x) \gets \top$ (see the following section on \MTL{} semantics).
A \emph{program}
is a finite set of safe rules; 
it is forward- or backwards-propagating  if so are all its rules.
 
An expression  is \emph{ground} if it mentions no variables.
A \emph{fact} is
an expression of the form $\matA @ \varrho$ with $\matA$ a
ground relational atom and $\varrho$ an interval; a
\emph{dataset}  is a finite set of facts.
The \emph{coalescing}
of facts $M@\varrho_1$ and $M@\varrho_2$ with   union-compatible
intervals $\varrho_1$ and $\varrho_2$
is the fact $M@(\varrho_1 \cup \varrho_2)$. 
The \emph{coalescing}  of a dataset $\D$, denoted by $\coal{\D}$, is the unique dataset obtained by iteratively coalescing facts in $\D$ until no more facts can be coalesced.
The \emph{grounding} 
$\ground{\Prog}{\D}$ 
of a program $\Prog$ with respect to a dataset $\D$ is the set of all ground rules that can be obtained 
by assigning constants in $\Prog$ or $\D$  to variables in  $\Prog$.

The \emph{dependency graph} of a  program $\Prog$ is the
directed graph $G_{\Prog}$,
with a vertex $v_P$ for each predicate $P$ in $\Prog$
and an edge $(v_Q, v_R)$ whenever there is a rule in $\Prog$ mentioning
$Q$ in the body and $R$ in the head.
Program $\Prog$ is \emph{recursive} if
$G_{\Prog}$ has a cycle.
A predicate $P$ is \emph{recursive} in   $\Prog$
if $G_{\Prog}$ has 
a path ending in $v_P$ and including
a cycle (the path can be a self-loop); otherwise, it is \emph{non-recursive}.
A metric atom is \emph{non-recursive} in $\Prog$ if so are 
all its predicates; otherwise it is \emph{recursive}.
The (non-)recursive fragment of $\Prog$ is the (unique) subset of
rules in $\Prog$ with (non-)recursive atoms in heads.
For a predicate $P$, a rule  $r$ is \emph{$P$-relevant} 
in $\Prog$ if 
there exists a rule $r'$ in $\Prog$ mentioning $P$ or $\bot$ in the
head and a path in $G_{\Prog}$ starting from 
a vertex representing the predicate in the head of $r$ and ending in a vertex representing some predicate 
from the  body  of $r'$.

\subsection{Semantics and Reasoning Problems}\label{sec:semantics}

An \emph{interpretation} $\I$ is a function which assigns to each time point $t $
a set of ground relational atoms;
if  an atom $P(\cbf)$ belongs to this set,  we say that $P(\cbf)$  is \emph{satisfied} at $t$ in $\I$ and we write ${\I,t \models P(\cbf)}$.
This extends to arbitrary metric atoms as shown in Table~\ref{semantics}.
The meaning of past operators $\diamondminus$, $\boxminus$, and $\Si$ is additionally visualised in figures below.
The first figure shows that $\diamondminus_{[1,2]} P$ holds at $3.1$ if $P$ holds at some time point located within the interval $[1.1, 2.1]$.
The second figure indicates that $\boxminus_{[1,2]} P$ holds at $3.1$ if $P$ holds continuously in the interval $[1.1, 2.1]$.
The third figure, in turn, shows that $Q \Si_{[1,2]} P$ holds at $3.1$ if $P$ holds at some time point within $[1.1, 2.1]$ and $Q$ holds continuously from this time point up to the time point $3.1$.
The meaning of future operators $\diamondplus$, $\boxplus$, and $\Ui$  is symmetric to the meaning of the corresponding past operators $\diamondminus$, $\boxminus$, and $\Si$, respectively.
\begin{table}[t]
	\begin{alignat*}{3}
		&\I, t \models  \top    && && \text{for each } t
		\\
		&\I, t  \models \bot   && && \text{for no } t
		\\
		&\I,  t  \models \diamondminus_\varrho M    && \text{iff}   && \I, t' \models  M \text{ for some } t' \text{ with } t -   t' \in \varrho
		\\
		&\I,  t  \models  \diamondplus_\varrho  M  && \text{iff} &&  \I,  t' \models    M  \text{ for some } t'  \text{ with } t' - t \in \varrho
		\\
		&\I, t  \models \boxminus_\varrho  M  && \text{iff} && \I, t' \models M \text{ for all } t' \text{ with } t-t' \in \varrho
		\\
		&\I,t  \models \boxplus_\varrho  M   && \text{iff} && \I,t' \models  M \text{ for all } t'  \text{ with } t'-t \in \varrho
		\\
		&\I,t  \models M_1 \Si_\varrho M_2   &&  \text{iff} &&  \I,t' \models  M_2 \text{ for some } t'  \text{ with } t-t' \in \varrho \text{ and }  \I,t'' \models  M_1 \text{ for all } t'' \in (t',t)
		\\
		&\I,t  \models  M_1  \Ui_\varrho M_2 \qquad   &&  \text{iff} \qquad && \I, t' \models  M_2 \text{ for some } t'  \text{ with } t' - t \in \varrho \text{ and }
		\I, t'' \models  M_1 \text{ for all } t'' \in (t,t')
	\end{alignat*}
	\caption{Semantics of ground metric atoms}
	\label{semantics}
\end{table}

\begin{center}
\begin{tikzpicture}[scale = 0.25]
	
\tikzset{>=latex}
	
\draw[->, thick, gray](0,0)--(13,0);
	
\foreach \i in {1,3,5,9}
{
	\draw[-, thick] (\i,0.4) -- (\i , -0.4);
}		
\node[ label={[yshift=-23]{$1.1$}}] at (1,0) {};
\node[ label={[yshift=-23]{$2.1$}}] at (5,0) {};

\node[ label={[yshift=2]{$ P$}}] at (3,0) {};

\node[ label={[yshift=-23] $3.1$}] at (9,0) {};
\node[ label={[yshift=2]{$ \diamondminus_{[1,2]}  P$}}] at (9,0) {};
%

\node[ label={[yshift=-23]{$\mathbb{Q}$}}] at (12.8,0) {};
\end{tikzpicture}
\qquad 
\begin{tikzpicture}[scale = 0.25]
\tikzset{>=latex}
	
\draw[->, thick, gray](0,0)--(13,0);
	
\foreach \i in {1,5,9}
{
	\draw[-,  thick] (\i,0.4) -- (\i , -0.4);
}		
\node[ label={[yshift=-23]{$1.1$}}] at (1,0) {};
\node[ label={[yshift=-23]{$2.1$}}] at (5,0) {};


\node[ label={[yshift=-23] $3.1$}] at (9,0) {};
\node[ label={[yshift=2]{$ \boxminus_{[1,2]}  P$}}] at (9,0) {};
%
	\draw[-,  thick, decoration={brace, raise=7pt,amplitude=5pt}, decorate] (1,0) -- node[yshift=19.5pt, align=center]  {$P$}
	(5,0);

\node[ label={[yshift=-23]{$\mathbb{Q}$}}] at (12.8,0) {};
\end{tikzpicture}
\qquad 
\begin{tikzpicture}[scale = 0.25]	
\tikzset{>=latex}
	
\draw[->, thick, gray](0,0)--(13,0);
	
\foreach \i in {1,3,5,9}
{
	\draw[-, thick] (\i,0.4) -- (\i , -0.4);
}		
\node[ label={[yshift=-23]{$1.1$}}] at (1,0) {};
\node[ label={[yshift=-23]{$2.1$}}] at (5,0) {};

\node[ label={[yshift=4]{$ P$}}] at (2.6,0) {};

\node[ label={[yshift=-23] $3.1$}] at (9,0) {};
\node[ label={[yshift=2]{$ Q \mathcal{S}_{[1,2]}  P$}}] at (11.5,0) {};
%
	\draw[-,  thick, decoration={brace, raise=7pt,amplitude=5pt}, decorate] (3,0) -- node[yshift=19.5pt, align=center]  {$Q$}
	(9,0);

\node[ label={[yshift=-23]{$\mathbb{Q}$}}] at (12.8,0) {};

\end{tikzpicture}
\end{center}

An interpretation $\I$ satisfies a  fact  $\matA @ \varrho$
if $\I,t \models \matA$ for all $t \in \varrho$. Interpretation $\I$ satisfies a ground rule $r$ if, whenever $\I$ satisfies each body atom of $r$ at a time point $t$, then $\I$ also satisfies the head of $r$ at $t$.
Interpretation $\I$ satisfies a (non-ground) rule $r$ if it satisfies each ground  instance of~$r$. Interpretation $\I$ is a \emph{model} of a program $\Prog$ if
it satisfies each rule in $\Prog$, and it is a \emph{model} of a dataset $\D$ if
it satisfies each fact in $\D$. Each dataset $\D$ has a unique least model $\I_\D$, and we say that dataset $\D$ \emph{represents}  interpretation $\I_{\D}$. 
Program $\Prog$ and dataset $\D$
are \emph{consistent} if they have a model, and they
\emph{entail} a fact $M@ \varrho$
if each  model of both $\Prog$ and $\D$ is  a model of $M@ \varrho$.

\emph{Consistency checking} is the problem of checking whether a given program  and  a
dataset  admit a common model.
 \emph{Fact entailment} is the problem of checking
whether a program and a dataset entail a given relational fact.
Consistency checking and fact entailment in $\MTL$  reduce to the complements of each other.
Specifically, to check whether a program $\Prog$ and a dataset $\D$ are inconsistent, it suffices to check whether they entail
fact $P@0$, for $P$ a fresh 
proposition occurring neither in $\Prog$ nor in $\D$.
In turn, to check whether $\Prog$ and $\D$ entail a fact $P(\cbf)@\varrho$, it suffices to 
check whether the following program and dataset
are inconsistent, with $P'$ a fresh predicate
of the same arity as $P$, $\xbf$ a tuple of distinct variables, and
$t$ an arbitrary time point belonging to the interval $\varrho$:
\begin{align*}
\Prog'' = \Prog \cup \{\bot \gets P'(\xbf) \land \boxminus_{\varrho_1} P(\xbf) \land \boxplus_{\varrho_2} P(\xbf)  \},  
&&
\D''   =  \D  \cup \{P'(\cbf) @ t \}.
\end{align*}
Intervals 
 $\varrho_1$ and $\varrho_2$ are constructed using $\varrho$ and $t$;
for example, 
if $\varrho =[ t_1,t_2 )$, then
$\varrho_1=[0,t-t_1]$ and $\varrho_2=[0,t_2-t)$, whereas if $t_2= \infty$, then $t_2-t$ stands for $\infty$.

\subsection{Fixpoint Characterisation and the Canonical Interpretation}\label{sec:fixpoint}

Each pair of a consistent program $\Prog$ and a dataset $\D$ admits a
unique least model,
which we refer to as  their \emph{canonical interpretation} $\can{\Prog}{\D}$  \cite{brandt2018querying}.
As in plain Datalog, we can construct this interpretation by applying rules of $\Prog$ to $\D$
in a forward-chaining manner until a fixpoint is reached. 
Unlike plain Datalog, however,  the fixpoint  in \MTL{} may only be reachable
after infinitely many materialisation steps; 
for example, given a fact $P@0$, the rule $\boxplus_1 P \gets P$ propagates $P$ to all positive integers, which requires $\omega$ materialisation steps.

Materialisation is performed using the \emph{immediate consequence operator} $T_{\Prog}$;
intuitively, $T_{\Prog}(\I)$ can be seen as the interpretation obtained from $\I$
by adding all relational facts that can be derived by applying once 
all rules in
$\Prog$ which are not $\bot$-rules in all possible ways to $\I$.
Formally, we define   $T_{\Prog}$, for a program $\Prog$,
as the operator mapping each interpretation $\I$  to the least interpretation  containing
$\I$ and satisfying the following property for each
ground instance $r$ of a rule in $\Prog$ that is not a $\bot$-rule: whenever 
$\I$ satisfies each body atom of $r$ at a time point $t$, then
$T_{\Prog}(\I)$ satisfies the head of $r$ at $t$.
Subsequent applications of  $T_{\Prog}$ to the (least) model $\I_\D$ of $\D$  define the following sequence of interpretations, for all ordinals~$\alpha$:
\begin{align*}
T_{\Prog}^0(\I_\D) &= \I_\D ,&&
\\
T_{\Prog}^{\alpha}(\I_\D) & = T_{\Prog} \left( T_{\Prog}^{\alpha-1}(\I_\D) \right),  && \text{for } \alpha \text{ a successor ordinal},
\\
T_{\Prog}^{\alpha} (\I_\D) &= \bigcup_{\beta < \alpha} T_{\Prog}^{\beta}(\I_\D),  &&  \text{for }
\alpha \text{ a limit ordinal}.
\end{align*}
The canonical interpretation is obtained after at most $\omega_1$ 
(i.e., the smallest ordinal number that, considered as a set, is uncountable) applications of $T_{\Prog}$;
that is,   $\can{\Prog}{\D} = T_{\Prog}^{\omega_1}(\I_\D)$   \cite{brandt2017ontology}; in turn,
if $\Prog$ and $\D$ are consistent then $\can{\Prog}{\D}$ is the least
common model of $\Prog$ and $\D$.
Furthermore, $\bot$-rules can be treated as constraints, and so  $\Prog$ and $\D$ are consistent if and only if $\can{\Prog}{\D}$ is a model of all $\bot$-rules in $\Prog$.


Although the timeline in \MTL{}  is dense, it can be divided into regularly distributed intervals
which are uniform in the sense that, in the canonical interpretation, the same relational atoms hold in all time points belonging to the same interval.
This observation was first exploited
to partition the rational timeline,
for a program $\Prog$ and dataset $\D$,  into
punctual intervals $[i\cdot d,i\cdot d]$ and open intervals $((i-1)\cdot d, i\cdot d)$, for
each $i \in \Z$, where $d$ is the greatest common divisor (gcd) of the numbers occurring 
as interval endpoints in  $\Prog$ and  $\D$ \cite{brandt2018querying}.
Later, an alternative partitioning of the timeline was
proposed~\cite{walega2019datalogmtl}, where
punctual intervals of the form $[i\cdot d,i\cdot d]$ are replaced with punctual intervals 
${
[t + 
i \cdot d', t +
i \cdot d']}$, for all rational numbers $t$  in $\D$,
${i \in \mathbb{Z}}$,
and 
$d'$ the gcd of numbers occurring in $\Prog$; 
in turn, open intervals of the form
$((i-1)\cdot d, i\cdot d)$ were replaced with open intervals located between the new punctual intervals.
Below we present example  partitionings of the timeline into intervals stemming from both discretisation methods, for the case 
where the only rationals occurring in $\D$ are $\frac{1}{2}$ and $\frac{2}{3}$ and the gcd of $\Prog$ is 1 (therefore, $d = \frac{1}{6}$ and $d'=1$).
The second partitioning  has the advantage
that the gcd is computed independently from $\D$; this was exploited to
devise reasoning techniques with a better computational behaviour in data size \cite{walega2019datalogmtl}. 

\medskip
\begin{center}
	\begin{tikzpicture}[scale=1.15]
		\foreach \x in {0,1,...,3} {
		\node[draw=none,fill=none] (l\x) at (\x*3,0.2) { $\x$};
		}
	
		\foreach \x in {1,2} {
		\node[draw=none,fill=none] (l\x12) at (\x*3+1.5,0.2) { $\x\!\frac{1}{2}$};
		\node[draw=none,fill=none] (l\x23) at (\x*3+2,0.2) { $\x\!\frac{2}{3}$};
		}
	
		\node[draw=none,fill=none] (l012) at (1.5,0.2) { $\frac{1}{2}$};
		\node[draw=none,fill=none] (l023) at (2,0.2) { $\frac{2}{3}$};
		
		\foreach \x in {0,1,2} {
				\node[circle,inner sep=1pt,draw=black,fill=black,outer sep=1pt] (p\x) at (\x*3,.6) {};
				\node[circle,inner sep=1pt,draw=black,fill=black,outer sep=1pt] (pp\x) at (\x*3+1.5,.6) {};
				\node[circle,inner sep=1pt,draw=black,fill=black, outer sep=1pt] (ppp\x) at (\x*3+2,.6) {};
			}
		\node[circle,inner sep=1pt,draw=black,fill=black,outer sep=1pt] (p3) at (9,.6) {};
	
	\foreach \x in {0,1,2} {
		\draw[thick,(-)] (p\x) -- (pp\x);
		\draw[thick,(-)] (pp\x) -- (ppp\x);
	}
	\draw[dotted,thick] (-1.2,.6) -- (-.5,.6);
	\draw[thick,-)] (-.5,.6) -- (p0);
	\draw[thick,(-)] (ppp0) -- (p1);
	\draw[thick,(-)] (ppp1) -- (p2);
	\draw[thick,(-)] (ppp2) -- (p3);
	\draw[thick,(-] (p3) -- (9.5,.6);
	\draw[thick,dotted] (9.5,.6) -- (10.2,.6);
	
	\foreach \x in {0,1,...,9} {
		\node[circle,inner sep=1pt,draw=black,fill=black,outer sep=1pt] (z\x) at (\x,2) {};
	}
	\foreach \x in {0,1,...,8} {
		\node[circle,inner sep=1pt,draw=black,fill=black,outer sep=1pt] (zz\x) at (\x+0.5,2) {};
	}
	\foreach \x in {0,1,...,8} {
		\draw[thick,(-)] (z\x) -- (zz\x);
	}
	\draw[dotted,thick] (-1.2,2) -- (-.5,2);
	\draw[thick,-)] (-.5,2) -- (z0);
	\draw[thick,(-)] (zz0) -- (z1);
	\draw[thick,(-)] (zz1) -- (z2);
	\draw[thick,(-)] (zz2) -- (z3);
	\draw[thick,(-)] (zz3) -- (z4);
	\draw[thick,(-)] (zz4) -- (z5);
	\draw[thick,(-)] (zz5) -- (z6);
	\draw[thick,(-)] (zz6) -- (z7);
	\draw[thick,(-)] (zz7) -- (z8);
	\draw[thick,(-)] (zz8) -- (z9);
	\draw[thick,(-] (z9) -- (9.5,2);
	\draw[thick,dotted] (9.5,2) -- (10.2,2);
	
	\foreach \x in {1,...,3} {
		\node[draw=none,fill=none] (k\x) at (\x*3,1.6) { $\x$};
	}

	\foreach \x in {1,2} {
		\node[draw=none,fill=none] (kk\x) at (\x*3+0.5,1.6) { $\x\!\frac{1}{6}$};
		\node[draw=none,fill=none] (kkk\x) at (\x*3+1,1.6) { $\x\!\frac{2}{6}$};
		\node[draw=none,fill=none] (kkkk\x) at (\x*3+1.5,1.6) { $\x\!\frac{3}{6}$};
		\node[draw=none,fill=none] (kkkkk\x) at (\x*3+2,1.6) { $\x\!\frac{4}{6}$};
		\node[draw=none,fill=none] (kkkkkk\x) at (\x*3+2.5,1.6) { $\x\!\frac{5}{6}$};
	}

\node[draw=none,fill=none] (0) at (0,1.6) { $0$};
\node[draw=none,fill=none] (16) at (0.5,1.6) { $\frac{1}{6}$};
\node[draw=none,fill=none] (13) at (1,1.6) { $\frac{2}{6}$};
\node[draw=none,fill=none] (12) at (1.5,1.6) { $\frac{3}{6}$};
\node[draw=none,fill=none] (23) at (2,1.6) { $\frac{4}{6}$};
\node[draw=none,fill=none] (56) at (2.5,1.6) { $\frac{5}{6}$};
	\end{tikzpicture}
\end{center}

%% file: sections/techniques.tex
\section{Reasoning Techniques for DatalogMTL}\label{sec:background}

In this section we recapitulate  different 
techniques available for reasoning in \MTL{}, which we  take as a starting point for the
development of our approach. In Section \ref{sec:naive},
we present a variant of the na\"ive
materialisation procedure  \cite{walkega2023finite,walega2021finitely} where the $k$-th iteration of its main loop, for any $k \in \mathbb{N}$, yields
a dataset representing the the $k$-th application of the 
immediate consequence operator of the input program to the input dataset.
In Section \ref{sec:LTL-translation}, we discuss the reduction of
\MTL{} reasoning to LTL satisfiability proposed by \citeN{brandt2018querying}. Finally, in Section \ref{sec:automata} we
describe the  automata-based  procedure exploited by \citeN{walega2019datalogmtl} to establish
optimal complexity bounds for reasoning.

\subsection{Na\"ive Materialisation}\label{sec:naive}

The fixpoint characterisation from Section \ref{sec:fixpoint}
suggests the
na\"ive materialisation-based  approach 
specified in Procedure~\ref{alg::naive}.
For a program $\Prog$, a dataset $\D$, and a  fact $P(\cbf)@\varrho$, the procedure 
applies the immediate consequence operator $T_\Prog$ to $\I_\D$ (Lines \ref{alg_n}, \ref{alg_coal}, and \ref{alg_passon})
until one of the following holds:
\begin{itemize}
\item[--] 
the partial materialisation of $\Prog$ and $\D$ yields inconsistency due to satisfaction of the body of some $\bot$-rule, in which case
 $\inc$ is returned (Line~\ref{incons}),
 
\item[--] 
the partial materialisation entails the input fact $P(\cbf)@\varrho$, in which case the truth value $\true$ and the obtained partial materialisation are returned (Line~\ref{derived}),
or 

\item[--] 
the application of $T_\Prog$  reaches a fixpoint and the obtained materialisation does not satisfy  $P(\cbf)@\varrho$ nor yields an inconsistency, in which case
the  value $\false$ and the full  materialisation are returned (Line~\ref{alg_passon}).
\end{itemize}

\begin{algorithm}[t]
\SetKwFor{Loop}{loop}{}{}
\SetKwInput{Input}{Input}
\SetKwInput{Output}{Output}
\SetKwComment{Comment}{$//$ }{}
\Input{A program $\Prog$, a dataset $\D$, and a fact $P(\cbf)@\varrho$}
\Output{Either $\inc$, or  a pair of a truth value and a dataset}

Initialise $\mathcal{N}$ to $\emptyset$ and $\D'$ to $\D$;\label{naive_init}

\Loop{}{\label{naiveline2}
	$\mathcal{N} \coloneqq \Prog[\D']$\Comment*{derive new facts} \label{alg_n} 
	$\mathcal{C} \coloneqq \coal{ \D' \cup \mathcal{N}}$\Comment*{add new facts to partial materialisation} \label{alg_coal}
	\textbf{if} the least model of $\D'$ does not satisfy some $\bot$-rule in $\Prog$ \textbf{then} \Return $\inc$\Comment*{inconsistency is derived}\label{incons}		
	\textbf{if} $\D' \models P(\cbf)@\varrho$ \textbf{then} \Return $(\true, \D')$\Comment*{input fact is derived}\label{derived}
	\textbf{if} $\mathcal{C} = \D'$ \textbf{then} \Return $(\false, \D')$\Comment*{fixpoint is reached}\label{alg_fixp}	
	$\D' \coloneqq \mathcal{C};$ \label{alg_passon}
}
\caption{Na\"ive($\Prog, \D, P(\cbf)@\varrho$)}
\label{alg::naive}
\end{algorithm}

\begin{example}\label{run_ex}
As a running example, consider the input fact $R_1(c_1,c_2)@[4,4]$, the dataset 
$$
\D_{\mathsf{ex}}=\{R_1(c_1,c_2)@[0,1],
\qquad
R_2(c_1, c_2)@[1,2],
\qquad
R_3(c_2,c_3)@[2,3],
\qquad  
R_5(c_2)@[0,1] \},
$$
and the program $\Prog_{\mathsf{ex}}$ consisting of the 
following rules:
\begin{align} 
	R_1(x,y) &\gets \diamondminus_{[1,1]} R_1(x,y)    \subtag{$r_1$} ,\\ 
	\boxplus_{[1,1]}R_5(y)&\gets R_2(x,y) \land \boxplus_{[1,2]}R_3(y,z) \subtag{$r_2$}, \\
	R_4(x)&\gets \diamondminus_{[0,1]}R_5(x) \subtag{$r_3$}, \\
	R_6(y)&\gets R_1(x,y) \land \boxminus_{[0,2]}R_4(y) \land R_5(y)  .\subtag{$r_4$}
\end{align}
\end{example}
In na\"ive materialisation, rules are applied by  first identifying the facts that can ground the rule body, and then determining the maximal intervals for which all the ground body atoms hold simultaneously. 
For instance, 
Rule $r_2$ in Example \ref{run_ex} is applied to $\D_{\mathsf{ex}}$
by matching 
$R_2(c_1, c_2)$ and $R_3(c_2,c_3)$ 
to the rule's body atoms,
and determining that $[1,1]$ is the maximal interval in which  the body atoms $R_2(c_1, c_2)$ and 
$\boxplus_{[1,2]} R_3(c_2,c_3)$ hold together.
As a result, the head
$\boxplus_{[1,1]} R_5(c_2)@[1,1]$ holds, and so the fact  $ R_5(c_2)@[2,2]$ is derived.
We formalise it below.

\begin{definition}\label{def::inst}
	Let $r$ be a  rule of the form 
	$M'  \leftarrow M_1 \land \dots \land M_n$ 
	for some  $n \geq 1$, which is not a $\bot$-rule, 
	and let $\D$ be a dataset.
	\noindent The set of \emph{instances} for $r$ and $\D$ is defined  as follows:
	\begin{multline*}
		\ins{r}{\D}  =  \big\{ ( M_1\sigma@\varrho_1 , \dots , M_n \sigma@\varrho_n ) \mid  \sigma \text{ is a substitution and, for each }   i \in \{1, \dots , n\} \\  
		\varrho_i \text{ is a subset-maximal interval such that  } \D \models M_i\sigma @ \varrho_i \big\}.
	\end{multline*}   
	The set $\der{r}{\D}$ of facts \emph{derived} by $r$ from $\D$ is defined as follows:
	\begin{multline}
		\der{r}{\D} = \{M \sigma@ \varrho \mid \sigma \text{ is a substitution, } M \text{ is the  relational atom in } M' \sigma,
		\text{ and }
		\\ 
	    \text{there is }  ( M_1 \sigma@\varrho_1 , \dots , M_n \sigma@\varrho_n) \in \ins{r}{\D} 
		\text{ such that } \varrho \text{ is the unique}
		\\ 
	 	\text{subset-maximal interval satisfying }  M' \sigma@(\varrho_1 \cap \ldots \cap \varrho_n) \models M\sigma@ \varrho \}.
		\label{eq:consequences}
	\end{multline}
	The set of facts \emph{derived} from $\D$ by one-step application of  $\Prog$  is
	\begin{equation}\label{eq:prog-cons}
		\Prog[\D] = \bigcup_{r \in \Prog} \der{r}{\D}.
	\end{equation}   
\end{definition}

Notice that, in the definition above, to determine instances for rules and facts derived by rules,
one needs to compute subset-maximal intervals consisting of time points satisfying a particular property.
Such computations are  straight forward, for example, if a particular property holds in the time point $1$, in all time points between $2$ (inclusive) and $3$ (exclusive), and for all time points bigger than $5.5$, then the subset-maximal intervals consisting of time points satisfying this property are $[1,1]$, $[2,3)$, and $(5, \infty)$.

By Definition \ref{def::inst}, 
the set $\mathcal{N}$ of facts derived by Procedure~\ref{alg::naive} 
in Line~\ref{alg_n}
in the first iteration of the loop on our running example is
$\Prog_{\mathsf{ex}}[\D_{\mathsf{ex}}] =
 \{R_1(c_1,c_2)@[1,2],  R_4(c_2)@[0,2],  R_5(c_2)@[2,2]\}.$ 
The partial materialisation $\D_{\mathsf{ex}}^1$ that will be passed to the next materialisation step is
obtained in Line~\ref{alg_coal} as $\D_{\mathsf{ex}}^1 = \coal{ \D_{\mathsf{ex}} \cup \Prog_{\mathsf{ex}}[\D_{\mathsf{ex}}]}$
where, as we described in Section~\ref{sec:syntax},  $\coal{\D}$ is the unique dataset obtained by exhaustively coalescing facts in $\D$. 
In our example,  $R_1(c_1,c_2)@[0,1]$ in and $R_1(c_1,c_2)@[1,2]$ will be coalesced into $R_1(c_1,c_2)@[0,2]$. Thus, 
\begin{multline*}
\D_{\mathsf{ex}}^1 = \{R_1(c_1,c_2)@[0,2], 
\quad
R_2(c_1, c_2)@[1,2], 
\quad
R_3(c_2,c_3)@[2,3],  
\quad
\\
R_4(c_2)@[0,2], 
\quad
R_5(c_2)@[0,1], 
\quad
R_5(c_2)@[2,2]\}.
\end{multline*}
In the second round, rules are applied to $\D_{\mathsf{ex}}^1$.
The application of $r_1$ derives 
$R_1(c_1,c_2)@[1,3]$ (from $R_1(c_1, c_2)@[0,2]$) and the application of $r_2$ rederives a redundant fact $R_5(c_2)@[2,2]$. 
In contrast to the previous step, Rule $r_4$ can now be applied to derive 
the new fact $R_6(c_2)@[2,2]$. 
Finally, $r_3$ derives the new fact $R_4(c_2)@[2,3]$ and rederives $R_4(c_2)@[0,2]$.
After coalescing, the second step of materialisation yields the 
following partial materialisation:
\begin{multline*}
\D_{\mathsf{ex}}^2 = \{R_1(c_1,c_2)@[0,3], 
\quad
R_2(c_1, c_2)@[1,2], 
\quad
R_3(c_2,c_3)@[2,3],  
\\
R_4(c_2)@[0,3], 
\quad
R_5(c_2)@[0,1], 
\quad
R_5(c_2)@[2,2],
\quad
R_6(c_2)@[2,2] \}.
\end{multline*}
In the third materialisation step,
rules are applied to $\D_{\mathsf{ex}}^2$, and derive
the new fact $R_1(c_1,c_2)@[1,4]$ using Rule  $r_1$, as well
as all facts which were already derived in the second materialisation step (with the only exception of $R_1(c_1,c_2)@[1,3]$).
After coalescing we obtain:
\begin{multline*}
\D_{\mathsf{ex}}^3 = \{R_1(c_1,c_2)@[0,4], 
\quad
R_2(c_1, c_2)@[1,2], 
\quad
R_3(c_2,c_3)@[2,3],  
\\
R_4(c_2)@[0,3], 
\quad
R_5(c_2)@[0,1], 
\quad
R_5(c_2)@[2,2],
\quad
R_6(c_2)@[2,2] \}.
\end{multline*}
At this point, the algorithm detects that $\D_{\mathsf{ex}}^3$ entails the input fact $R_1(c_1,c_2)@[4,4]$ and stops.

Procedure \ref{alg::naive} is both sound and complete. 
To show  this,
for each  $k \in \N$,  let $\mathcal{N}_k$ and $\D_k$ denote the contents of, respectively, $\mathcal{N}$ and $\D$ in Procedure~\ref{alg::naive}  upon the completion of the $k$th iteration of the loop. 
Then we prove soundness by showing inductively on $k\in \N$, that $\I_{\D_k} \subseteq T^k_\Prog(\I_\D)$. 
\begin{theorem}[Soundness]\label{soundness_n}
Consider Procedure \ref{alg::naive} running on input  $\Prog$ and $\D$. 
Upon the completion of the $k$th (for some $k 
\in \N$) iteration of the loop of  Procedure~\ref{alg::naive}, 
it holds that  $\I_{\D'} \subseteq  T_{\Prog}^{k}(\I_\D)$.
\end{theorem}
By Theorem~\ref{soundness_n}, if the least model $\I_{\D'}$ of $\D'$
 is not
a model of a $\bot$-rule in $\Prog$, then the canonical interpretation
$\can{\Prog}{\D}$ is also not a model of such a rule; thus, if the algorithm
returns $\inc$, then $\Prog$ and $\D$ are inconsistent.
Next,  to show completeness, we  prove 
inductively  that $T_\Prog^k(\I_\D) \subseteq \I_{\D_k}$. 
\begin{theorem}[Completeness]\label{complete_n}
Consider Procedure \ref{alg::naive} running on input  $\Prog$ and $\D$. For each $k \in \N$, upon the completion of the $k$th iteration of the loop of  Procedure~\ref{alg::naive},
it holds that  $  T_{\Prog}^{k}(\I_\D) \subseteq \I_{\D'} $.
\end{theorem}
If $\Prog$ and $\D$ are inconsistent,  the canonical intepretation
$\can{\Prog}{\D}$ is not a model of some $\bot$-rule $r$ in $\Prog$; thus,
there is an ordinal $\alpha$ such that $T_{\Prog}^{\alpha}(\I_\D)$ is also not a model of $r$,
 in which case
Theorem~\ref{complete_n} ensures that the algorithm returns
$\inc$, as required.

Furthermore, if a fixpoint is reached without encountering an inconsistency, then
our procedure ensures that the input fact is not entailed
and also that we have obtained 
a representation
 of the full canonical interpretation which can 
be kept in memory and subsequently exploited
for checking entailment of any other fact.
The procedure is, however, not always terminating as reaching a fixpoint may require infinitely
many rounds of rule applications.


\subsection{Translation to LTL}\label{sec:LTL-translation}

The discretisation of the rational timeline described in Section~\ref{sec:fixpoint}
was exploited by \citeN{brandt2018querying} to reduce reasoning in \MTL{} to reasoning in LTL.
The reduction transforms a program $\Prog$ and a dataset $\D$ into an LTL formula $\varphi_{\Prog,\D}$
such that 
$\Prog$ and $\D$ are consistent if and only if $\varphi_{\Prog,\D}$ is LTL-satisfiable.
Here, LTL is a propositional modal logic interpreted over the integer time points and equipped
with  temporal operators $\bigcirc_P$ for \emph{at the previous time point},
 $\Box_P$ for  \emph{always in the past},
$\Si$ for \emph{since}, 
 $\bigcirc_F$ for \emph{at the next time point},  $\Box_F$ for \emph{always in the future},
and $\Ui$ for \emph{until}. 
An LTL formula $\varphi$ is \emph{satisfiable} if it holds at  time point $0$ in some LTL model.

Since, in contrast to \MTL, the language of LTL is propositional, the first step in the translation is to ground $\Prog$ with all constants occurring in $\Prog$ or $\D$.
Then, every relational atom $P(\mathbf{c})$ occurring in the 
grounding of $\Prog$ with constants from $\Prog$ and $\D$ is translated into a propositional symbol $P^\mathbf{c}$.
Moreover,  the MTL operators occurring in  $\Prog$ are
rewritten using LTL operators.
Both the grounding of the input program $\Prog$ and the translation of the MTL operators (with binary-encoded numbers) into sequences of LTL operators yield an exponential blow-up.
For instance, 
assume that  $\Prog$ mentions an atom $\boxminus_{[0,60)} A(x)$,
both  $\Prog$ and $\D$ mention constants $c_1, \dots, c_{n}$, and 
that interval $(0,60)$ contains $m$ intervals after discretising the
timeline.
Then $\boxplus_{(0,60)} A(x)$ is  translated to an LTL-formula containing $n$ 
conjuncts (one conjunct for each $c_i $), each of the form
$
\bigcirc_F A^{c_i} \land \bigcirc_F \bigcirc_F A^{c_i} \land  \dots \land \underbrace{\bigcirc_F \dots \bigcirc_F }_{m} A^{c_i}.
$ 
Consequently, 
$\varphi_{\Prog,\D}$ is exponentially large.
Since satisfiability checking in LTL is \PS{}-complete, this approach provides a (worst-case optimal)
\EXPS{} reasoning procedure for \MTL{}.

\subsection{Automata-based Reasoning}\label{sec:automata}

An alternative approach to exploit the time discretisation  of \MTL{}
is to directly apply 
automata-based techniques, without the need of constructing an LTL-formula.
Such automata-based constructions are well-studied in the context of LTL, where checking satisfiability of a formula $\varphi$ reduces to checking non-emptiness of
a generalised non-deterministic B\"uchi automaton
where states are sets of formulas relevant to $\varphi$, the alphabet consists of sets of propositions, and the
transition relation and accepting conditions ensure that words accepted by the automaton are exactly the 
models of $\varphi$~\cite{baier2008principles}.
 
This technique can be adapted to the \MTL{} setting 
\cite{walega2019datalogmtl};
each state now represents ground metric atoms
holding at all time points within a fragment of the timeline called a \emph{window}. 
Since the timeline can be discretised in \MTL{}, each window can be finitely represented as a sequence consisting of sets of metric atoms which hold in the consecutive intervals from the window.
Additionally, for such a sequence to be a state of the automaton, it is required that the involved metric atoms are \emph{locally consistent}; for example,
if $\boxplus_{[0,\infty)}A$---which states that $A$ holds always in the future---holds in some interval $\varrho$, then $\boxplus_{[0,\infty)}A$ needs to hold also in all the intervals in the window which 
are to the right of $\varrho$ in the timeline.
The remaining components  
of the automaton are defined analogously to the
case of LTL.

More specifically, given  \MTL{} program $\Prog$ and dataset $\D$, the approach consists in 
constructing two generalised B\"uchi automata, which are responsible for recognising fragments of models of $\Prog$ and $\D$ that are located on the timeline to the left and to the right of  $\D$, respectively.
Thus, if both of these automata have non-empty languages, there exists a common model of $\Prog$ and $\D$, that is, $\Prog$ and $\D$ are consistent.
These automata have the same states, alphabet, and initial states, but they differ on the transition function and accepting conditions (the latter two components are symmetric for the left- and right-automata).
In particular, states $Q$ of these automata are windows (of the length computed based on the numbers occurring in the temporal operators of $\Prog$) and alphabet $\Sigma$ consists of all sets of metric atoms which are relevant for $\Prog$ and $\D$ (i.e., use predicates and constants occurring in $\Prog$ and $\D$).
The transition function 
$\delta$ is a partial function from $Q \times \Sigma$ to $Q$.
For the right-moving automaton $\delta$ takes as an input a current window $W$ and an alphabet symbol $\sigma$ (i.e. a set of metric atoms), and returns the next window obtained by shifting $W$ by one interval on the discritised timeline to the right (see \Cref{sec:fixpoint} for the description of time discretisatrion) and 
putting in this new interval all atoms from $\sigma$.
For the left-moving automaton, $\delta$ is defined similarly but the new windows is shifted to the left.
Accepting conditions for the right-moving automaton capture the meaning of `future' operators $\boxplus$ and $\Ui$,
whereas for the left-moving automaton the accepting conditions capture the  meaning of `past' operators $\boxminus$ and $\Si$.
In particular, if $\boxplus_\varrho A$ occurs in some window, the automaton needs to guarantee that both $A$ and $\boxplus_\varrho A$ will occur infinitely often successive states in the run.
If $A \Ui_\varrho B$ occurs in a window, the automaton needs to guarantee that $B$ will occur in future (in distance bounded by $\varrho$) and that $A$ occurs everywhere up to this occurrence of this $B$.

To sum up, consistency checking in \MTL{} reduces to checking non-emptiness of (pairs of) automata.
Importantly, this reduction provides an optimal \PS{}~upper bound  for data complexity \cite{walega2019datalogmtl}.
In particular, automata states are polynomially large in the size of the dataset since
 windows  can be chosen so that
the number of intervals in each window is 
polynomially large.
Moreover, the number of ground metric atoms that can hold in each of these intervals is also polynomially bounded.
Thus, each state is polynomially representable and   non-emptiness of the automata can be checked with the standard `on-the-fly' approach~\cite{baier2008principles} in \PS{}. 

We have implemented the automata-based approach for \MTL{} \cite{wang2022meteor}  using existing libraries in the Python 3.8 eco-system without depending on other third-party libraries.
The procedure takes as an input a program $\Prog$ together with a dataset $\D$ and returns a truth value indicating whether $\Prog$ and $\D$ are consistent or not.
We will call this procedure $\mathsf{AutomataProcedure}$ and use it in \Cref{alg:pipline} as a module of our reasoning system.

%% file: sections/practical.tex
\section{Our Practical Reasoning Algorithm} \label{sec:overview}

In this section we propose a scalable approach for deciding
both consistency and
fact entailment in \MTL{}.
The key elements of our approach are as follows:

\begin{itemize}
\item[--] an optimised 
materialisation procedure using a semina\"ive strategy 
which efficiently applies  the immediate 
consequence operator while minimising repeated inferences,

\item[--] a  realisation of the automata-based reasoning approach, and

\item[--]  an effective way of combining materialisation with automata-based reasoning
 that aims at  
resorting to materialisation-based reasoning
whenever possible.

\end{itemize}
In the remainder of this section we discuss each of the key components of
our approach in detail.
For convenience of presentation, and without loss of generality, we will
assume that input programs do not contain rules whose body is vacuously satisfied (i.e., satisfied  in the empty interpretation).
Observe that each such  rule can be formulated as a fact and added to the input dataset instead; for example, rule $P \gets \top$ is equivalent 
to a fact $P@(-\infty, \infty)$.

\subsection{Semina\"ive Materialisation}\label{subsec:seminaive}

In this section we present a semina\"ive materialisation procedure for \MTL{}.
Analogously to the classical semina\"ive algorithm for plain Datalog~\cite{abiteboul1995foundations},  
the main idea behind our procedure is to keep track of newly derived facts in each materialisation step by storing them in a set $\Delta$, and
to make sure that each rule application in the following materialisation step involve at least one fact in $\Delta$. 
In this way, the procedure will consider each  instance (for a rule and dataset, as we introduced in Definition~\ref{def::inst}) \emph{at most once} throughout its entire execution, and so, such a procedure is said to enjoy the \emph{non-repetition} property. 
The same fact, however, can still be derived multiple times by \emph{different}  instances; this type of redundancy is difficult to prevent and is not addressed by the standard semina\"ive strategy. 
 
Our aim  is to lift the semina\"ive rule evaluation strategy to the setting of \MTL{}.
As we have seen in Section~\ref{sec:naive}, a rule instance can be considered multiple times in the na\"ive materialisation procedure; for example, the  instance 
$(\diamondminus_{[0,1]}R_5(c_2)@[0,2])$
of Rule $r_3$ in Example \ref{run_ex}  is considered by the 
na\"ive materialisation procedure in Section~\ref{sec:naive} 
both in the first and second materialisation steps to derive  $R_4(c_2)@[0,2]$  since the na\"ive procedure cannot detect that the
fact $R_5(c_2)@[0,1]$  used to instantiate $r_3$ in the second step had previously 
been used to instantiate $r_2$.
Preventing such redundant computations, however, involves certain challenges.
First, by including in $\Delta$ just newly derived facts as in Datalog, we may
overlook information obtained by coalescing newly derived facts with previously derived ones.
Second, restricting application to relevant rule instances requires 
consideration of the semantics of metric operators in rule bodies.

Procedure~\ref{alg::seminaive} extends the semina\"ive strategy to the setting of \MTL{} while overcoming the aforementioned difficulties.
Analogously to the na\"ive approach, each iteration of the main loop captures a single materialisation step consisting of a round of rule applications (Line~\ref{alg2_5}) followed by the coalescing of relevant facts (Line~\ref{alg2_coal}); as before, dataset $\D'$ stores the partial materialisation resulting from each iteration and is initialised as the input dataset, whereas the dataset $\mathcal{N}$ stores the facts obtained as a result of rule application and is initialised as empty.
Following the rationale behind the semina\"ive strategy for Datalog,
newly derived information in each materialisation step is now stored as a dataset $\Delta$, which is initialised as the input dataset $\D$ and which is suitably maintained in each iteration (Line~\ref{alg2_delta}); furthermore,
Procedure \ref{alg::seminaive} ensures in Line \ref{alg2_5} that only rule instances
for which it is essential to involve facts from $\Delta$ (as formalised in the following definition) are taken into account during rule application. 

\begin{algorithm}[t]
	\SetKwFor{Loop}{loop}{}{}
	\SetKwInput{Input}{Input}
	\SetKwInput{Output}{Output}
	\Input{A program $\Prog$, a dataset $\D$, and a fact $P(\cbf)@\varrho$}
	\Output{Either $\inc$, or  a pair of a truth value and a dataset}
	
	Initialise $\mathcal{N}$ to $\emptyset$, and both $\Delta$ and $\D'$
	to $\D$\label{alg2_init};
	 
	\Loop{}{\label{line2}
		
		$\mathcal{N} \coloneqq \semi{\Prog}{\D'}{\Delta}$; \label{alg2_5}
		
		$\mathcal{C} \coloneqq \coal{\D' \cup \mathcal{N}}$; \label{alg2_coal}

		$\Delta \coloneqq \{ M@\varrho\in  \mathcal{C} \mid M@\varrho \text{ entails some fact  in } \mathcal{N} \text{  which is not entailed by } \D'\}; $\label{alg2_delta}	
	
	   	\textbf{if} the least model of $\D'$ does not satisfy some $\bot$-rule in $\Prog$ \textbf{then} \Return $\inc$; \label{incons-semi}	
	
	   	\textbf{if} $\D' \models P(\cbf)@\varrho$ \textbf{then} \Return $(\true, \D')$; \label{derived-semi}
	
		\textbf{if} $\Delta = \emptyset$ \textbf{then} \Return $(\false, \D')$; \label{alg2_7}
		
		$\D' \coloneqq \mathcal{C}$\label{alg2_8};
	}
	\caption{Semina\"ive($\Prog,\D,P(\cbf)@\varrho$)}
	\label{alg::seminaive}
\end{algorithm}

\begin{definition}\label{def::semi-inst}
Let $r$ be a rule of the form
$M'  \leftarrow M_1 \land \dots \land M_n$,
for some  $n \geq 1$,
and let  $\D$ and $\Delta$ be datasets.
The set of instances for $r$ and $\D$ relative to $\Delta$
is defined as follows:
$$
		\ins{r}{\D \threedotcolon \Delta}  =  \big\{  ( M_1\sigma@\varrho_1 , \dots , M_n \sigma@\varrho_n ) \in \ins{r}{\D}  \mid 
		\D \setminus \Delta \not\models
		M_i\sigma @\varrho_i ,  \text{ for some } i \in \{1, \dots, n \}  \big\}. \label{eq:rel-instance-delta}
$$
The set $\der{r}{\D \threedotcolon  \Delta}$ of facts derived by $r$ from $\D$ relative to $\Delta$ is defined analogously to $\der{r}{\D}$ in Definition \ref{def::inst}, with the exception that  $\ins{r}{\D}$ is replaced with $\ins{r}{\D \threedotcolon \Delta}$ in Expression \eqref{eq:consequences}.
Finally, the set $\semi{\Prog}{\D}{\Delta}$ of facts derived from $\D$
by one-step semina\"ive application of $\Prog$ is defined as $\Prog[\D]$ in Expression \eqref{eq:prog-cons}, by replacing  
$\der{r}{\D}$ with $\der{r}{\D \threedotcolon \Delta}$.
\end{definition}

In each materialisation step, Procedure~\ref{alg::seminaive} exploits  Definition~\ref{def::semi-inst} to identify as relevant those
rule instances with some `new' element (i.e., one that  
cannot be entailed 
without the facts in $\Delta$). 
The facts derived by such relevant rule instances in each iteration are stored in set $\mathcal{N}$ (Line~\ref{alg2_5}).
Note that all facts stored in sets $\mathcal{C}$, $\mathcal{N}$, and $\mathcal{D}'$ are relational and hence the entailment
checks needed to compute the set $\Delta$ can be
performed syntactically by checking containment between intervals; for instance
a relational fact $R_4(c_2)@[0,2]$ entails
fact $R_4(c_2)@[0,1]$ because of the inclusion $[0,1] \subseteq [0,2]$, but it does not entail the
fact $R_4(c_2)@[1,3]$ since $[1,3] \not\subseteq [0,2]$.

As in the na\"ive approach, rule application is followed by a coalescing step where the partial materialisation is updated with the facts derived from rule application (Line \ref{alg2_coal}). In contrast to the na\"ive approach, however, Procedure~\ref{alg::seminaive} needs to maintain set $\Delta$ to ensure that it captures only 
new facts.  This is achieved in  Line \ref{alg2_delta},
where a fact in the updated partial materialisation
is considered new if it entails a fact in $\mathcal{N}$ that was not already entailed by the
previous partial materialisation.
The procedure terminates 
in Line~\ref{incons-semi} if an 
inconsistency has been derived,
in
Line~\ref{derived-semi} if the input fact has been derived, or in Line~\ref{alg2_7} if  $\Delta$ is empty. Otherwise, the procedure carries over the updated partial materialisation and the set of newly derived facts to the next materialisation step.

We next illustrate the application of the procedure to $\D_{\mathsf{ex}}$ and $\Prog_{\mathsf{ex}}$ and our example fact $R_1(c_1,c_2)@4$
 from Example~\ref{run_ex}.
In the first materialisation step, all input facts are considered as newly derived (i.e., $\Delta = \D$) and hence 
$\mathcal{N} = \semi{\Prog}{\D'}{\Delta} = \Prog[\D']$ and the result of coalescing coincides with the partial materialisation computed by the na\"ive procedure (i.e., $\mathcal{C} = \D_{\mathsf{ex}}^1)$.
Then, the procedure identifies as new all facts in $\D_{\mathsf{ex}}^1$ which are not entailed by $\D_\mathsf{ex}$, namely:
$$
\Delta = \{
R_1(c_1,c_2)@[0,2], 
\quad
R_4(c_2)@[0,2], 
\quad
R_5(c_2)@[2,2]
 \}
 .
$$
In the second step, rule evaluation is performed  in Line~\ref{alg2_5};
since $\Delta$ is used for this,
the procedure no longer considers
the redundant instance  $( R_2(c_1,c_2)@[0,2], \boxplus_{[1,2]}R_3(c_2,c_3)@[1,2] )$
of $r_2$,
 since $\boxplus_{[1,2]}R_3(c_2,c_3)@[1,2]$ is  entailed already by $\D_{\mathsf{ex}}^1 \setminus \Delta$.
Similarly, 
the redundant instance 
$(\diamondminus_{[0,1]}R_5(c_2)@[0,2])$
of Rule $r_3$  is not considered,
as 
$\D_{\mathsf{ex}}^1 \setminus \Delta \models \diamondminus_{[0,1]}R_5(c_2)@[0,2]$.
In contrast, all non-redundant facts derived by the na\"ive strategy are also derived by the semina\"ive procedure and after coalescing dataset $\mathcal{C} = \D_{\mathsf{ex}}^2$. 
The set $\Delta$ is now updated as follows: 
$$
\Delta = \{ R_1(c_2,c_2)@[0,3],
\qquad
R_4(c_2)@[0,3],
\qquad  
R_6(c_2)@[2,2]\}. 
$$
In particular, note that $\Delta$ contains the coalesced fact
$R_4(c_2)@[0,3]$ instead of fact $R_4(c_2)@[2,3]$ derived from rule application. 
Datasets $\Delta$ and $\D' = \D_{\mathsf{ex}}^2$ are passed on to the third materialisation step, where all redundant computations identified 
in Section~\ref{sec:naive} are avoided with the only exception 
of fact $R_6(c_2)@[2,2]$, which is re-derived using the instance of $r_4$ consisting of  $R_1(c_2, c_2)@[0,3]$,   $\boxminus_{[0,2]}R_4(c_2)@[2,3]$, and $R_5(c_2)@[2,2]$. 
Note that
this is a new instance which was not used in previous iterations, and hence it does not violate  the non-repetition property.
 Note also that, 
as with the na\"ive strategy, our semina\"ive procedure terminates on our running example
in this materialisation step since input fact $R_1(c_1,c_2)@[4,4]$ has been derived.

We conclude this section by establishing correctness of our 
procedure, similarly as we did in 
Theorems~\ref{soundness_n} and \ref{complete_n}
for the  na\"ive strategy from Procedure~\ref{alg::naive}.

\begin{theorem}[Soundness]\label{soundness_sm}
	Consider Procedure \ref{alg::seminaive} running on input  $\Prog$ and $\D$. 
	Upon the completion of the $k$th (for some $k 
	\in \N$) iteration of the loop of  Procedure~\ref{alg::seminaive}, 
	it holds that  $\I_{\D'} \subseteq  T_{\Prog}^{k}(\I_\D)$.
\end{theorem}
Completeness is proved by induction on the number $k$ of iterations of the main loop.
In particular, we show that if $  T_{\Prog}^{k+1}(\I_\D)$ satisfies a new fact $M@t$, then there must be a rule $r$ and an instance in $\ins{r}{\D_k \threedotcolon \Delta_k}$ witnessing the derivation of $M@t$.
Note that since
$\ins{r}{\D_k \threedotcolon \Delta_k} \subseteq \ins{r}{\D_k }$,
 it requires strengthening the argument from the proof of Theorem~\ref{complete_n}, where it was sufficient to show 
 that such an instance is in  $\ins{r}{\D_k }$.
This will imply that each fact satisfied by $  T_{\Prog}^{k+1}(\I_\D)$ is derived in the $k+1$st iteration of our procedure.

\begin{theorem}[Completeness]\label{complete_sm}
	Consider Procedure \ref{alg::seminaive} with input program $\Prog$ and input dataset $\D$. For each $k \in \N$, upon the completion of the $k$th iteration of the loop of  Procedure~\ref{alg::seminaive},
	it holds that  $  T_{\Prog}^{k}(\I_\D) \subseteq \I_{\D'} $.
\end{theorem}

\subsection{Optimised Semina\"ive Evaluation
}\label{subsec:oseminaive}

Although the semina\"ive procedure enjoys the non-repetition property, it
can still re-derive facts already obtained in previous materialisation steps; as a result,
it can incur in a potentially large number of redundant computations. In particular, 
 as discussed in Section \ref{subsec:seminaive}, fact $R_6(c_2)@[2,2]$ is re-derived using Rule $r_4$ in the third materialisation 
 step of our running example, and  (if the procedure did not terminate) it would also be re-derived in all subsequent materialisation steps (by different instances of Rule $r_4$).

In this section, we present an optimised variant of our semina\"ive procedure which 
further reduces the number of redundant computations performed 
during materialisation. The main idea is to disregard rules during 
the execution of the procedure as soon as we can be certain 
that their application will never derive new facts in subsequent 
materialisation steps. In our example, Rule~$r_4$ can be discarded
after the second materialisation step as its application will only 
continue to re-derive  fact $R_6(c_2)@[2,2]$ in each subsequent materialisation step.

To this end, we will exploit the distinction between recursive and non-recursive predicates in a program, as defined
in  Section~\ref{sec:Preliminaries}.
In the case of our Example~\ref{run_ex},
predicates $R_2$, $R_3$, $R_4$, and $R_5$ are non-recursive, whereas $R_1$ and $R_6$ are recursive, since in the dependency graph of the program $\Prog_{\mathsf{ex}}$---depicted below---only vertices corresponding to $R_1$ and $R_6$ can be reached by paths containing a cycle.
Hence, Rules $r_2$ and $r_3$ are non-recursive in $\Prog_{\mathsf{ex}}$ since their heads mention non-recursive predicates, whereas $r_1$ and $r_4$ are recursive.
\begin{center}
\begin{tikzpicture}
[scale = 0.5,>=stealth']

\tikzset{>=latex}
				
\node[ label={[yshift=-13,xshift=-7]{$v_{R_1}$}}](a1) at (0,0) {\textbullet} ;

\node[ label={[yshift=-13,xshift=-7]{$v_{R_2}$}}](a2) at (0,-1) {\textbullet} ;

\node[ label={[yshift=-13,xshift=-7]{$v_{R_3}$}}](a3) at (0,-2) {\textbullet} ;

\node[ label={[yshift=-21,xshift=2]{$v_{R_4}$}}](a4) at (4,-1.8) {\textbullet} ;

\node[ label={[yshift=-5,xshift=2]{$v_{R_5}$}}](a5) at (4,-0.5) {\textbullet} ;

\node[ label={[yshift=-21,xshift=2]{$v_{R_6}$}}](a6) at (9,-0.5) {\textbullet} ;

\draw[->] (a1) to [out=130+90,in=320+90+90,looseness=8] (a1);
\node at (7,1.5) {} ;

\path[->] (a2) edge [bend left=20]  node[pos=0.65,above]() {$ $} (a5);
\path[->] (a3) edge [bend right=20]  node[pos=0.65,above]() {$ $} (a5);

\path[->] (a4) edge [bend right=0]  node[pos=0.65,above]() {$ $} (a5);

\path[->] (a1) edge [bend left=20]  node[pos=0.65,above]() {$ $} (a6);
\path[->] (a5) edge [bend right=0]  node[pos=0.65,above]() {$ $} (a6);
\path[->] (a4) edge [bend right=20]  node[pos=0.65,above]() {$ $} (a6);

\end{tikzpicture}
\end{center}
Now, in contrast to recursive predicates, for which new facts can be derived in each materialisation step, the materialisation of 
non-recursive predicates will be completed after linearly many materialisation steps;
from then on, the rules will no longer derive any new facts involving these predicates.
This observation can be exploited to optimise our semina\"ive evaluation as follows. 
Assume that the procedure has fully materialised all non-recursive predicates in the input program. 
At this point, we can safely discard
all non-recursive rules;  furthermore, we can also discard a recursive rule $r$ with a non-recursive body atom $M$ if the current partial materialisation does not entail any grounding of $M$ (in this case, $r$ cannot apply in further materialisation steps).
An additional optimisation applies to forward-propagating 
programs, where rules cannot propagate information `backwards' along the timeline; in this case, we can compute the 
maximal time points  for which each non-recursive body atom in $r$ may possibly hold, select the minimum $t_r$ amongst such values, and discard $r$ as soon as we can determine that the materialisation up to time point $t_r$ has been fully completed.

In the case of our Example~\ref{run_ex},  materialisation of the non-recursive predicates $R_2$, $R_3$, $R_4$, and $R_5$  is complete after two materialisation steps.
Hence, at this point we can disregard non-recursive Rules $r_2$ and $r_3$ and focus on the recursive forward-propagating Rules $r_1$ and $r_4$. Furthermore, the maximum time point at which the non-recursive predicates $R_4$ and $R_5$ from $r_4$ can hold are $3$ and $2$, respectively, and hence $t_{r_4} = 2$; as upon the completion of the second materialisation step we can be certain that the remaining body atom of $r_6$, namely $R_1$, has been materialised up to $t_{r_6}$,  we can  discard the rule $r_6$.
In subsequent materialisation steps
we can apply only Rule $r_1$, thus avoiding many redundant computations.

Procedure~\ref{alg::optseminaive} implements these ideas by extending our  semina\"ive materialisation from Lines~\ref{line:n}--\ref{line:false}. 
In particular, in each materialisation step, the procedure checks whether all non-recursive predicates have been fully materialised (Line~\ref{same}), 
in which case it removes all non-recursive rules from the input program as well as all recursive rules with an unsatisfied non-recursive body atom (Lines~\ref{delprognr}--\ref{empty}). 
It also  sets the flag to $1$, which activates the additional optimisation for forward propagating programs, which is applied in Lines~\ref{forward}--\ref{proj}.

\begin{algorithm}[t]
	\SetKwFor{Loop}{loop}{}{}
	\SetKwInput{Input}{Input}
	\SetKwInput{Output}{Output}
	\Input{A program $\Prog$, a dataset $\D$, and a fact $P(\cbf)@\varrho$}
	\Output{Either $\inc$, or  a pair of a truth value and a dataset}
	
	Initialise $\mathcal{N}$ to $\emptyset$, 
	$\Delta$ and $\D'$ to $\D$, $\Prog'$ to $\Prog$,  
	$flag$ to $0$, and $S_r$ (for each $r \in \Prog$) to the set of body atoms in $r$ that are non-recursive in $\Prog$;
	
	\Loop{}{\label{loopstart}
		
		$\mathcal{N} \coloneqq \semi{\Prog'}{\D'}{\Delta}$; \label{line:n}
		
		$\mathcal{C} \coloneqq \coal{\D' \cup \mathcal{N}}$; \label{line:c}	

		$\Delta \coloneqq \{ M@\varrho\in  \mathcal{C} \mid M@\varrho \text{ entails some fact  in } \mathcal{N} \text{ but which is not entailed by } \D'\}; $\label{alg_delta}	

		 	\textbf{if} the least model of $\D'$ does not satisfy some $\bot$-rule in $\Prog$ \textbf{then} \Return $\inc$; \label{incons-semi-opt}
		 			
			\textbf{if} $\D' \models P(\cbf)@\varrho$ \textbf{then} \Return $(\true,  \D')$; \label{derived-semi-opt}
	
			\textbf{if} $\Delta = \emptyset$ \textbf{then} \Return $(\false,  \D')$;\label{line:false}	
		
		\If{$flag = 0$ and  the datasets $\D'$ and $\mathcal{C}$ satisfy the same facts with non-recursive in $\Prog'$ predicates\label{same}}
		{
			Set $flag$ to $1$ and $\Prog'$ to the recursive fragment of $\Prog$; \label{delprognr}
			
			\For{each $r \in \Prog'$}{
				\lIf{there is $M \in S_r$ such that $\D' \not\models M\sigma@ t$, for each substitution $\sigma$ and time point $t$  } { $\Prog' \coloneqq \Prog' \setminus \{r\}$} \label{empty}
			}
		} 
		
		\If{$flag = 1$ and $\Prog'$ is forward propagating\label{forward}} 
		{
			
			\For{each $r \in \Prog'$}{
			\For{each $M \in S_r$}{
				$t_{max}^M \coloneqq$ maximum right endpoint amongst all intervals $\varrho$ satisfying $\D' \models M\sigma@\varrho$, for some  substitution $\sigma$;
			}
				
				$t_{r} \coloneqq $  minimum value in $\{t_{max}^M \mid M \in S_r\}$;
				
				\lIf{
					for each  $M@\varrho \in \mathcal{C}$ with $\varrho$ overlapping $(-\infty, t_r]$, there is  $M@\varrho' \in \D'$ with $\varrho \cap (-\infty,t_r] \subseteq \varrho'$
					\label{proj}} {$\Prog' \coloneqq \Prog' \setminus \{r\}$}
			}
			
		}
		$\D' \coloneqq \mathcal{C}$;\label{loopend}
	}
	\caption{OptimisedSemina\"ive($\Prog,\D, P(\cbf)@\varrho)$}
	\label{alg::optseminaive}
\end{algorithm}

We conclude this section by establishing correctness of our procedure. 
We first observe that, as soon as the algorithm switches the flag to $1$, we can be certain that all facts with  non-recursive predicates have been fully materialised.

\begin{lemma}\label{non-rec}
Consider Procedure~\ref{alg::optseminaive} running on input program $\Prog$ and dataset $\D$.
Whenever  $flag = 1$, dataset $\D'$ satisfies all facts over a 
 non-recursive predicate in $\Prog$ that are 
entailed by $\Prog$ and $\D$.
\end{lemma} 
This lemma shows us that once the $flag$ is changed to 1, Procedure~\ref{alg::optseminaive} can safely ignore all the non-recursive rules, as they can no longer be used to derive any new facts.
Thus, such rules are deleted from $\Prog$ in 
Line~\ref{empty}.
The deletion of rules in Line~\ref{proj}, on the other hand,
is based on a simple observation that a derivation of a new fact $M@t$ by a forward-propagating program is based only on facts holding at time points smaller-or-equal to $t$.
Thus, if for a forward-propagating program $\Prog'$ (condition in Line \ref{forward}) it holds that $\I_{\D'}$ and $T_{\Prog'}(\I_{\D'})$ satisfy the same facts within $(-\infty, t_r]$ (condition in Line \ref{proj}), 
then 
$\I_{\D'}$ and $T^\alpha_{\Prog'}(\I_{\D'})$ will coincide over $(-\infty, t_r]$, for each ordinal number $\alpha$.
By the construction, $t_r$ is the maximum time point in which the body of rule $r$ can hold, so if the above conditions are satisfied, we can safely delete  $r$ from the program in Line~\ref{proj}.
Consequently, we obtain the following result.

\begin{lemma}\label{lemma:invalid}
Whenever Procedure \ref{alg::optseminaive} removes a rule $r$ from $\Prog'$ in either
Line~\ref{empty} or~\ref{proj},  
$\can{\Prog'}{ \mathcal{C}} = \can{\Prog'\setminus 
\{r\}}{  \mathcal{C}}$.	
\end{lemma}
Finally, using Lemma~\ref{lemma:invalid} together with the soundness and completeness of our semina\"ive evaluation (established in Theorems~\ref{soundness_sm} and \ref{complete_sm}), we can show soundness and completeness of the optimised version of the procedure.

\begin{theorem}[Soundness and Completeness]\label{seminaive_theorem}
Consider Procedure \ref{alg::optseminaive} with input program $\Prog$ and input dataset $\D$. For each $k \in \N$, upon the completion of the $k$th iteration of the loop of  Procedure~\ref{alg::seminaive},
it holds that  $  T_{\Prog}^{k}(\I_\D) = \I_{\D'} $.	
\end{theorem}
Note that our optimisation for forward-propagating programs in Lines \ref{forward}--\ref{proj} can be adopted in a straightforward way for backwards-propagating programs, as these  cases are symmetric.

We conclude this section by recalling that Procedure \ref{alg::optseminaive} is non-terminating, as it is possible for the 
main loop to
run for an unbounded number of iterations. In our approach, we will also exploit
a terminating (but incomplete) variant of the procedure, which we call OptimisedSemina\"iveHalt.
This variant is obtained by adding an additional
termination condition just before Line \ref{loopend} which breaks the loop and returns
$\Prog'$ and $\D'$ (without a truth value) if the flag has been set to 1. 
Thus, such a variant of the procedure
ensures termination with a (partial) materialisation once all the non-recursive predicates in the input program
have been fully materialised, and also returns a subset of relevant rules in the program.

\subsection{Combining Materialisation and Automata}\label{sec:combined}

We are now ready to provide a practical and scalable reasoning algorithm combining 
optimised semina\"ive materialisation and automata construction. Our algorithm 
delegates  the bulk of the computation to the materialisation
component and resorts to automata-based techniques only as needed to
ensure termination and completeness.

\SetAlgorithmName{Algorithm}{}{}

\begin{algorithm}[t]
	\SetKwInput{Input}{Input}
	\SetKwInput{Output}{Output}
	\SetKwFor{Loop}{loop}{}{}
	\SetKwFor{threadone}{thread 1:}{}{}
	\SetKwFor{threadtwo}{thread 2:}{}{}
	\Input{A program $\Prog$, a dataset $\D$, and a fact $P(\cbf)@\varrho$}
	\Output{Either $\inc$, or a pair  of a truth value  and a dataset}
	
	$\Prog' \coloneqq$ the set of   $P$-relevant  rules in $\Prog$; \label{relevantrules}
	
	$\D' \coloneqq$ the set of  facts in $\D$ which mention predicates occurring in $\Prog'$; \label{relevantdata}
	
	Out $\coloneqq$ OptimisedSemina\"iveHalt($\Prog',\D',P(\cbf)@\varrho)$; \label{nonrec}

     \lIf{Out $=(\Prog'',\D'')$}{assign $\Prog' \coloneqq \Prog''$ and $\D' \coloneqq \D''$}\label{alg_line}
	 \textbf{else} \textbf{return}  Out; \label{return-pre-mat}
	
    Run two threads in parallel:\\
		
		\textbf{Thread 1:} \\
		\qquad \textbf{return} OptimisedSemina\"ive($\Prog',\D',P(\cbf)@\varrho)$; \label{rec-materialisation}
	
		\textbf{Thread 2:} \\
			\qquad $\Prog'', \D'' \coloneqq \mathsf{EntailToInconsist}(\Prog', \D', P(\cbf)@\varrho)$; \label{ent-to-inc}  \\
			\qquad $b \coloneqq $  $\mathsf{AutomataProcedure}(\Prog'',\D'')$; \label{automata-proc} \\
			\qquad \textbf{return} the negation of $(b,\D')$; \label{return-automata}		   
	
	\caption{Practical reasoning algorithm}
	\label{alg:pipline}
\end{algorithm}

Our approach is summarised in Algorithm \ref{alg:pipline}.
When given as input a  program $\Prog$, a dataset $\D$, and a fact $P(\cbf)@\varrho$, the algorithm starts 
by identifying the subset $\Prog'$ of the input rules and the subset $\D'$ of the input data that are relevant to deriving the input fact $P(\cbf)@\varrho$ (Lines~\ref{relevantrules} and~\ref{relevantdata}).
This optimisation is based on the observation that a program
$\Prog$ and dataset $\D$ entail a fact of the form $P(\mathbf{c}) @ \varrho$ if and only if
so do the subset $\Prog^P$ of $P$-relevant rules in $\Prog$ and the subset
$\D^P$ of facts in $\D$ mentioning only predicates from $\Prog^P$.
Then, the algorithm proceeds according to the following steps. 
\begin{enumerate}
\item Pre-materialise $\Prog'$ and $\D'$ using the terminating (but possibly incomplete) version of the 
semina\"ive materialisation procedure OptimisedSemina\"iveHalt discussed at the end of  Section~\ref{subsec:oseminaive}. 
If the procedure terminates
because an inconsistency has been found, or the input fact has been derived, or a fixpoint has been reached, then
we can terminate and report  output (Line \ref{return-pre-mat}); otherwise, we have obtained a partial materialisation where
non-recursive predicates have been fully materialised and we can proceed to the next stage.

\item Run two threads in parallel. In the first (possibly non-terminating) thread, continue performing optimised semi-na\"ive materialisation until the 
input fact (or an inconsistency) is derived or a fixpoint is reached. In turn, in the second thread we 
resort to automata-based techniques applied to the partial 
materialisation obtained in the first stage of the algorithm; in particular, this second thread
applies in Line \ref{ent-to-inc} the  reduction from fact entailment to inconsistency checking as described in Section~\ref{sec:semantics}, and then
applies in Line \ref{automata-proc} the automata-based decision procedure of  \citeN{walega2019datalogmtl}, as described in Section~\ref{sec:automata}.
As soon as a thread returns a result, the other thread is terminated.
\end{enumerate}
It follows directly from the results proved earlier in this section and the theoretical results of
\citeN{walega2019datalogmtl} that 
Algorithm~\ref{alg:pipline} is sound, complete, and terminating.
Furthermore, the algorithm is designed so that, on the one hand,
materialisation is favoured over automata-based reasoning and,
on the other hand, the 
automata construction relies on a 
recursive program  that is as small as possible as well as on a partial materialisation
that is as complete as possible.
The favourable computational behaviour of this approach 
will be confirmed by our experiments.

%% file: sections/experiments.tex
\section{Implementation and Evaluation}\label{sec:experiments}

We have implemented our practical decision procedure from Algorithm \ref{alg:pipline} in a system
called MeTeoR. Our system is implemented in Python 3.8 and does not
depend on third-party libraries. 
MeTeoR is available for download and its demo can also be accessed via 
an online interface.\footnote{The version of MeToeR used in this paper is available at \url{https://github.com/wdimmy/DatalogMTL_Practical_Reasoning} and its online demo is available at \url{https://meteor.cs.ox.ac.uk/}.}
In this section, we describe the implementation choices made in MeTeoR and report the results of an extensive
empirical evaluation on available benchmarks.

\subsection{Implementation Details}\label{sec:impl-details}


A scalable implementation of materialisation-based reasoning requires 
suitable representation and storage schemes for facts.
In MeTeoR, we associate to each ground relational atom
a list of intervals sorted by their left endpoints, which
provides a compact
account of all facts mentioning a given relational atom.
Each ground relational atom is further indexed by a composite key consisting of its
predicate and its tuple of constants. This layout is useful
for fact coalescing
and fact entailment checking;  for instance,
to check if a fact  $M @ \varrho$ is entailed
by a dataset $\D$,
it suffices to find and interval $\varrho' $ such that
$M @ \varrho' \in \D$ and  $\varrho \subseteq \varrho'$; this can be achieved by
first scanning the sorted list of intervals for $M$
using the index and checking
if $\varrho$ is a subset of one of these intervals.
Additionally, each ground relational atom is also indexed by
each of its term arguments  to facilitate joins.
Finally, when a fact is inserted into the dataset,
the corresponding list of intervals  is sorted to
facilitate subsequent operations.

MeTeoR also implements  an optimised version of (temporal) joins, which are
required to evaluate rules with multiple body atoms.
A na\"ive implementation of the join of
metric atoms $M_1, \dots, M_n$ occurring in the body of a rule would require computing all intersections of intervals $\varrho_1, \dots, \varrho_n$ such that $M_i @ \varrho_i$ occurs in the so-far constructed dataset, for each $i \in \{1, \dots,  n\}$.
Since each $M_i$ may hold in multiple intervals in the dataset, the na\"ive approach is ineffective in practice.
In contrast, we implemented
a variant of the classical sort-merge join algorithm: we first sort all $n$  lists of intervals corresponding to $M_1, \dots, M_n$, and
then  scan the sorted lists to compute the intersections, which improves performance.
Our approach to temporal joins can  be seen as
 a  generalisation of the idea
sketched by \citeN[Section 5]{brandt2018querying}, which deals with two metric atoms at a time but has not
been put into practice to the best of our knowledge.

\subsection{Baselines and Machine Configuration}

We compared MeTeoR with two baselines: the query rewriting approach
by \citeN{brandt2018querying} and the reduction to LTL reasoning proposed by
\citeN{brandt2018querying} which was later implemented by \citeN{Yang-2022}. 
We could not compare our approach
to the temporal exension of the Vadalog system recently proposed by \citeN{tempVada}
as the system is not accessible.

The query rewriting approach by \citeN{brandt2018querying} is based
on rewriting a target predicate $P$ with respect to an input program $\Prog$
into an SQL query  that, when evaluated over the input dataset $\D$,
provides the set of all facts with maximal intervals
over $P$  entailed by $\Prog$ and $\D$.   The only implementation of this approach that we are aware of is the extension 
of the Ontop system described by \citeN{guzel2018ontop}; this implementation, however, is no longer
available\footnote{Personal communication with the authors.} and hence we have produced our
own implementation. 
Following~\citeN{brandt2018querying}, we used temporary tables (instead of subqueries)
to compute the extensions of predicates appearing in $\Prog$ on the fly, which has
two implications. First, we usually have not just one SQL query but a set of
queries for computing the final result; second, similarly to MeTeoR,
the approach
essentially works bottom-up rather than top-down.
The implementation by \citeN{brandt2018querying} provides two
variants for coalescing: the first one
uses standard SQL queries by
\citeN{zhou2006efficient}, whereas the second one implements coalescing
explicitly.
For our baseline we adopted the standard SQL approach, which
is less dependent on implementation details. Finally, we chose Postgresql 10.18 as the 
underpinning database for
all our baseline experiments.

As shown by \citeN{brandt2018querying}, \MTL{} reasoning can
be reduced to LTL reasoning by means of an exponential translation. 
This approach has been implemented by \citeN{Yang-2022} using
BLACK  \cite{DBLP:conf/tableaux/GeattiGM19} as the LTL reasoner of choice, and we used their implementation as a baseline
for our experiments.

All experiments have been conducted on a Dell PowerEdge R730 server with 512 GB RAM
and two Intel Xeon E5-2640 2.6 GHz processors running Fedora 33, kernel version 5.8.17. 
In each experiment, we verified that
the answers returned by MeTeoR and the baselines  coincide.

\subsection{Benchmarks}\label{sec:benchmarks}

The \emph{Temporal LUBM Benchmark} is an extension of
the Lehigh University Benchmark~\cite{guo2005lubm} with temporal rules and data, which we developed~\cite{wang2022meteor}.
LUBM's data generator, which provides means for generating datasets of different sizes, 
has been extended to randomly assign an interval to
each generated fact; the  rational endpoints of each interval belong to a  range, which is a  parameter.  
In addition, 
the plain Datalog rules  from the
OWL 2 RL fragment of LUBM have been extended with $29$  rules involving recursion and mentioning all
metric operators in \MTL{}. 
The resulting  \MTL{} program is denoted by~$\Prog_L$.

The \emph{Weather Benchmark} is based on  freely available
data with meteorological observations in the US \cite{maurer2002long}\footnote{\url{https://www.engr.scu.edu/~emaurer/gridded_obs/index_gridded_obs.html}};
in particular, we used a dataset $\D_W$ consisting of 397  million temporal facts spanning over the years 1949--2010, and then considered smaller subsets of this dataset.
Similar data was used in experiments of  \citeN{brandt2018querying}, however the format of data has slightly changed, so we adapted their (non-recursive) program 
(consisting of 4 rules)
for detecting
US states affected by excessive heat and heavy wind,
so that the program matches the data.
The obtained program is still non-recursive and, in what follows, it is denoted by $\Prog_W$.
 
The \emph{iTemporal Benchmark}  
provides a synthetic generator for \MTL{} programs and datasets \cite{iTemporal}.\footnote{\url{https://github.com/kglab-tuwien/iTemporal.git}}
The benchmark can generate different types of \MTL{} programs and equips them with  datasets. 
We used iTemporal to generate 10 recursive programs $\Prog_I^1, \dots , \Prog_I^{10} $ with different structures and
containing 20-30 rules each, as well as corresponding datasets of various size.




\subsection{Experiments}\label{sec:exper}

In this section, we describe  experiments that we performed for benchmarking MeTeoR.\footnote{All the datasets and programs that we used  in the  experiments are available together with exploited MeTeoR implementation at \url{https://github.com/wdimmy/DatalogMTL_Practical_Reasoning}.}

\paragraph{Automata vs LTL.}
Although MeTeoR utilises automata to ensure completeness and termination,   
we could have alternatively  relied on the translation to LTL
 implemented by \citeN{Yang-2022} together with
BLACK  \cite{DBLP:conf/tableaux/GeattiGM19} as the LTL reasoner of choice.
To determine which choice is favourable, 
we  considered
programs $\Prog_I^1, \dots , \Prog_I^{10}$ from the iTempora benchmark, together with  datasets
containing  $1,000$ facts each. 
For each pair of a dataset and a program, we  constructed  $10$ query facts; these facts were generated by 
first randomly choosing a predicate that  appears in a program or in a dataset; second,  constants are randomly chosen among those present in the program and in the dataset; third, an interval for the fact is generated. 
Next, we
performed fact entailment by first reducing to inconsistency checking and then either
applying the automata-based approach
or constructing  an LTL-formula and checking its satifiability with BLACK.
Average running times per 10 query facts  for both approaches are reported in Figure~\ref{fig:ltlautomata} (left),
where
each point represents the running times for the automata-based approach (vertical coordinate) and
LTL-based approach (horizontal coordinate) for a single program and dataset.
We can observe that the automata-based approach is consistently about an order of magnitude faster, which
justifies our choice over an LTL-based reasoner. 

\input{results/automata_ltl}

\paragraph{Scalability of automata.} We have conducted scalability experiments 
to understand the practical limitations of the automata-based approach. 
Based on the results from the previous experiment, we chose an `easy' program $\Prog_I^E$, a `medium' program
$\Prog_I^M$, and a `hard' program $\Prog_I^H$ amongst programs $\Prog_I^1, \dots, \Prog_I^{10}$, as depicted in Figure~\ref{fig:ltlautomata} (left).
We used iTemporal to generate increasing size datasets for these programs and we randomly generated query facts as in the previous experiment. 
The runtimes of automata-based approach for such inputs are  summarised in
Figure~\ref{fig:ltlautomata} (right),
 where we observe that the automata-based procedure is able to 
solve non-trivial problems, but struggles to scale beyond datasets with a few thousand temporal facts.


\paragraph{Comparison of materialisation strategies.}  We compared the na\"ive, semina\"ive, and optimised semina\"ive
materialisation procedures described in Sections \ref{sec:naive}, \ref{subsec:seminaive},
and \ref{subsec:oseminaive}, respectively. 
To this end, for each of the programs
$\Prog_i^{E}$, $\Prog_{i}^{M}$, and $\Prog_i^{H}$ 
we used iTemporal to generate a dataset containing one million facts. 
In
Figure~\ref{fig::matcomparison} we present running time and  memory consumption (measured as the 
maximum number of facts stored in memory) over the first 30 materialisation steps in each case.
Note that Theorems \ref{soundness_sm}, \ref{complete_sm}, and \ref{seminaive_theorem} ensure
that the na\"ive, semina\"ive, and optimised semina\"ive
materialisation procedures will compute
exactly the same output after any fixed number
of materialisation steps.
As we can see, the semina\"ive procedure consistently outperforms the
na\"ive approach  both in terms of running time and memory consumption, especially as materialisation progresses.
In turn, the optimised semina\"ive approach disregards rules after $6$th, $7$th, and $8$th
materialisation steps for $\Prog_I^{E}$,  $\Prog_I^{M}$, and $\Prog_I^{H}$, respectively; from then onwards,
the optimised procedure  outperforms the basic semina\"ive procedure. 
We can  note that after several steps of materialisation 
the memory usage stabilises;  the number of facts  in memory stops increasing but  their intervals keep growing.
This behaviour seems to be specific to programs we used in this experiment.

\input{results/mat}


\paragraph{Scalability of  optimised semina\"ive materialisation.} 
Our previous experiments provide evidence of the superiority of optimised
semina\"ive materialisation over the two alternative approaches.
We further evaluated the scalability of optimised
semina\"ive materialisation as the size of the data increases.
To this end, we considered the three iTemporal programs $\Prog_I^{E}$, $\Prog_I^{M}$, and $\Prog_I^{H}$
as well as program
$\Prog_L$ from the Temporal LUBM Benchmark, and
$\Prog_W$ from the Weather Benchmark.
In
Figure~\ref{fig::matscalability} we show running times and memory consumption
after the first $10$ materialisation steps for datasets with up to $10$ million temporal facts.
We note here that the program $\Prog_W$ is non-recursive and its full materialisation is obtained already after  $3$ materialisation steps; all other programs are recursive.
Our results suggest that the optimised semina\"ive
materialisation procedure exhibits favourable scalability behaviour as 
data size increases.

\input{results/scale_materialisation}


\paragraph{Comparison with query rewriting.} 
We compared the performance of MeTeoR with
the query rewriting baseline.  As the latter does not support recursive programs, we considered  non-recursive programs only.
For this, we generated non-recursive subsets of the program $\Prog_L$ from the Temporal LUBM Benchmark and two subsets of the program $\Prog_W$ from the Weather Benchmark.
In particular, we considered fragments  $\Prog_L^{1}$ (with $5$ rules), 
$\Prog_L^{2}$ ($10$ rules), and
$\Prog_L^{3}$ ($21$ rules)
 of $\Prog_L$
with rules relevant (see the end of Section~\ref{sec:syntax} for the precise definition of this notion) to 
predicates
$\mathit{ResearchAssistant}$, $\mathit{Lecturer}$, and $\mathit{FullProfessorCandidate}$, 
respectively.
For the Weather Benchmark we used fragments $\Prog_W^1$ and $\Prog_W^2$ (each with 2 rules) of $\Prog_W$, with rules relevant to predicates
$\mathit{HeatAffectedState}$ and $\mathit{HeavyWindAffectedState}$, respectively.
For each program we generated a collection of datasets with
$100,000$ facts each
and compared the runtime of MeTeoR with the baseline 
in the task of
computing all entailed facts over the   predicates mentioned above.
Note that since the programs are non-recursive, MeTeoR runs only (optimised semina\"ive) materialisation without exploiting automata.
Figure~\ref{fig:comparesql} presents results, where
each point represents the running times for
MeTeoR (vertical coordinate) 
and
 the baseline (horizontal coordinate) for particular pairs of a  program and a dataset;
 MeTeoR  consistently outperformed the baseline.

\input{results/compare_sql}

 
\paragraph{Usage of materialisation and automata in MeTeoR.} 
Our previous experiments provide strong indication that the
optimised semina\"ive materialisation algorithm implemented
in MeTeoR is more scalable than the automata-based component and that it can successfully deal
with complex rules and datasets containing millions of temporal facts.
Next, we assess whether Algorithm~\ref{alg:pipline} is indeed effective in
delegating most of the  
reasoning workload to the scalable materialisation component. 
To this end, we considered programs $\Prog_I^1, \dots, \Prog_I^{10}$ generated by iTemporal together with their corresponding datasets with 1000 facts each.
For each pair of a program and a dataset we randomly generated
100 query facts
and checked, for which facts entailment  can be checked by MeTeoR
using  materialisation
(i.e., Algorithm  \ref{alg:pipline} terminates in
either Line \ref{nonrec} or Line \ref{rec-materialisation}) and for which by the automata component (i.e., Algorithm  \ref{alg:pipline} terminates in
Line \ref{return-automata}).
In $99.7\%$ of the cases MeTeoR
decided entailment using materialisation only with
an average runtime of $0.17s$ and a standard deviation of $0.089s$. In turn, the automata component was
used in only 
$0.3\%$ cases, with an average runtime of $181.9$s and a
standard deviation of $0.089$s.

Although these results are obviously biased by the way iTemporal generates programs and datasets, they do support 
our hypothesis that focusing on the materialisation component in MeTeoR can provide a practically efficient reasoning mechanism.


%% file: results/automata_ltl.tex
\begin{figure}

\begin{tikzpicture}
\tikzstyle{every node}=[font=\scriptsize]
\begin{axis}[
xshift=0cm,
yshift=0cm,
width = 6cm,
height = 6cm,
xmin=0,xmax=2000,
ymin=0,ymax=2000,
axis line style={white},
tick style={white},
axis background/.style={fill=Blue4},
grid = both,
major grid style={white,thick},
minor grid style={draw=none},
ylabel={Run time (s) for automata},
ylabel style={yshift=-0.5em},
xlabel={Run time (s) for LTL},
xlabel style={yshift=0.5em},
legend style={at={(0.5,1.1)}, anchor=center,align=center},
legend columns=2,
]

\addplot[Light,
only marks, mark options={fill=none},
mark=triangle*, mark size=2.5pt]
coordinates {
(1342.3,219.2)(892.2,102.23)(783.2,98.2)(1862.1,298.3)(1312.3,193.2)(1233.1,234.1)(893.2,167.3)(1248.2,198.6)(1493.3,222.3)(749.3,86.3)
};

\addplot[Red1,
only marks, mark options={fill=none},
mark=triangle*, mark size=2.9pt]
coordinates {
(1862.1,298.3)
};

\addplot[Blue5,
mark=square*, line width=1pt, mark size=2.3pt]
coordinates {
(1233.1,234.1)
};

\addplot[Green3,
mark=otimes*, line width=1pt, mark size=2.3pt]
coordinates {
(749.3,86.3)
};

\node at (63,12) {$\Prog_I^E$};
\node at (113,32) {$\Prog_I^M$};
\node at (180,40) {$\Prog_I^H$};

\addplot [white, line width=1pt, samples at={0,2000}] {x};
\end{axis}

\begin{axis}[
xshift=7cm,
yshift=0cm,
width = 6cm,
height = 5.5cm,
xmin=1000, xmax=5000,
xtick={1000, 2000, 3000, 4000, 5000},
ymin = 0, ymax=5000,
ytick={0,1000, 2000, 3000, 4000, 5000},
axis line style={white},
tick style={white},
axis background/.style={fill=Blue4},
grid = both,
major grid style={white,thick},
minor grid style={draw=none},
ylabel={Run time~(s) for automata},
ylabel style={yshift=-0.5em},
xlabel={Data size},
xlabel style={yshift=0.5em},
legend style={at={(0.5,1.09)}, anchor=center,align=center},
legend columns=3,
]

\addplot[Green3, mark=otimes*, line width=1pt, mark size=1.2pt] table {results/new/automata_weather_scale.txt};
\addplot[Blue5, mark=square*, line width=1pt, mark size=1.2pt] table {results/new/automata_lubm_scale.txt};
\addplot[Red1,mark=triangle*, line width=1pt, mark size=1.2pt] table {results/new/automata_itemporal_scale.txt};

\legend{$\Prog_I^{E}$, $\Prog_I^{M}$, $\Prog_I^{H}$};
\end{axis}

\end{tikzpicture}		

%
%
%

\caption{Comparison of automata- and LTL-based approaches (left) and scalability of the automata approach (right)} 
\label{fig:ltlautomata}
\end{figure}
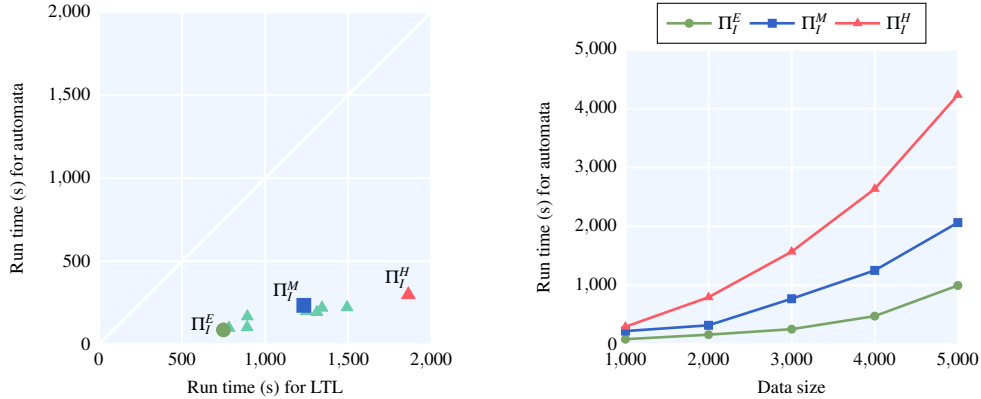

%% file: results/mat.tex
\begin{figure*}[ht]
\centering
\begin{tikzpicture}
\tikzstyle{every node}=[font=\scriptsize]
\begin{axis}[
xshift=0cm,
yshift=0cm,
width=0.33\textwidth,
height =4.5cm,
xmin=0, xmax=30,
ymin = 0, ymax=200,
axis line style={white},
tick style={white},
axis background/.style={fill=Blue4},
grid = both,
major grid style={white,thick},
minor grid style={draw=none},
ylabel=Run time~(s),
ylabel style={yshift=-0.5em},
xlabel={$\Prog_I^E$},
xlabel style={yshift=4cm},
legend style={at={(1.9,1.26)}, anchor=center,align=center},
legend columns=3,
]
\addlegendentry{Na\"ive \quad}
\addplot[Red1,line width=1pt, dash pattern=on 0.25cm off 0.05cm] table {results/new/mat_easy_time_naive.txt};
\addlegendentry{Semina\"ive \quad}
\addplot[Blue5, line width=1pt] table {results/new/mat_easy_time_semi.txt};
\addlegendentry{Optimised semina\"ive}
\addplot[Green3, line width=1pt, dash pattern=on 0.25cm off 0.05cm on 0.05cm off 0.05cm] table {results/new/mat_easy_time_opt.txt};
\end{axis}

\begin{axis}[
xshift=4cm,
yshift=0cm,
width=0.33\textwidth,
height =4.5cm,
xmin=0, xmax=30,
ymin = 0, ymax=200,
axis line style={white},
tick style={white},
axis background/.style={fill=Blue4},
grid = both,
major grid style={white,thick},
minor grid style={draw=none},
xlabel={$\Prog_I^M$},
xlabel style={yshift=4cm},
]
\addplot[Red1,line width=1pt, dash pattern=on 0.25cm off 0.05cm] table {results/new/mat_itemporal_time_naive.txt};
\addplot[Blue5, line width=1pt] table {results/new/mat_itemporal_time_semi.txt};
\addplot[Green3, line width=1pt, dash pattern=on 0.25cm off 0.05cm on 0.05cm off 0.05cm] table {results/new/mat_itemporal_time_opt.txt};
\end{axis}

\begin{axis}[
xshift=8cm,
yshift=0cm,
width=0.33\textwidth,
height =4.5cm,
xmin=0, xmax=30,
ymin = 0, ymax=200,
axis line style={white},
tick style={white},
axis background/.style={fill=Blue4},
grid = both,
major grid style={white,thick},
minor grid style={draw=none},
xlabel={$\Prog_I^H$},
xlabel style={yshift=4cm},
]
\addplot[Red1,line width=1pt, dash pattern=on 0.25cm off 0.05cm] table {results/new/mat_itemporal_hard_time_naive.txt};
\addplot[Blue5, line width=1pt] table {results/new/mat_itemporal_hard_time_semi.txt};
\addplot[Green3, line width=1pt, dash pattern=on 0.25cm off 0.05cm on 0.05cm off 0.05cm] table {results/new/mat_itemporal_hard_time_opt.txt};
\end{axis}

\begin{axis}[
xshift=0cm,
yshift=-3.5cm,
width=0.33\textwidth,
height =4.5cm,
xmin=0, xmax=30,
ymin = 100000, ymax=500000,
ytick = {100000, 200000, 300000, 400000, 500000},
scaled y ticks=false,
axis line style={white},
tick style={white},
axis background/.style={fill=Blue4},
grid = both,
major grid style={white,thick},
minor grid style={draw=none},
ylabel={Max. memory usage},
ylabel style={yshift=-0.5em},
]
\addplot[Red1,line width=1pt, dash pattern=on 0.25cm off 0.05cm] table {results/new/mat_lubm_memory_naive.txt};
\addplot[Blue5, line width=1pt] table {results/new/mat_lubm_memory_semi.txt};
\addplot[Green3, line width=1pt, dash pattern=on 0.25cm off 0.05cm on 0.05cm off 0.05cm] table {results/new/mat_lubm_memory_opt.txt};
\end{axis}

\begin{axis}[
xshift=4cm,
yshift=-3.5cm,
width=0.33\textwidth,
height =4.5cm,
xmin=0, xmax=30,
ymin = 100000, ymax=500000,
ytick = {100000, 200000, 300000, 400000, 500000},
scaled y ticks=false,
axis line style={white},
tick style={white},
axis background/.style={fill=Blue4},
grid = both,
major grid style={white,thick},
minor grid style={draw=none},
xlabel={Number of iterations},
]
\addplot[Red1,line width=1pt, dash pattern=on 0.25cm off 0.05cm] table {results/new/mat_itemporal_memory_naive.txt};
\addplot[Blue5, line width=1pt] table {results/new/mat_itemporal_memory_semi.txt};
\addplot[Green3, line width=1pt, dash pattern=on 0.25cm off 0.05cm on 0.05cm off 0.05cm] table {results/new/mat_itemporal_memory_opt.txt};
\end{axis}

\begin{axis}[
xshift=8cm,
yshift=-3.5cm,
width=0.33\textwidth,
height =4.5cm,
xmin=0, xmax=30,
ymin = 100000, ymax=500000,
ytick = {100000, 200000, 300000, 400000, 500000},
scaled y ticks=false,
axis line style={white},
tick style={white},
axis background/.style={fill=Blue4},
grid = both,
major grid style={white,thick},
minor grid style={draw=none},
]
\addplot[Red1,line width=1pt, dash pattern=on 0.25cm off 0.05cm] table {results/new/mat_itemporal_hard_memory_naive.txt};
\addplot[Blue5, line width=1pt] table {results/new/mat_itemporal_hard_memory_semi.txt};
\addplot[Green3, line width=1pt, dash pattern=on 0.25cm off 0.05cm on 0.05cm off 0.05cm] table {results/new/mat_itemporal_hard_memory_opt.txt};
\end{axis}
\end{tikzpicture} 
\caption{Comparison of time and memory consumption in various materialisation strategies}
\label{fig::matcomparison}	
\end{figure*}
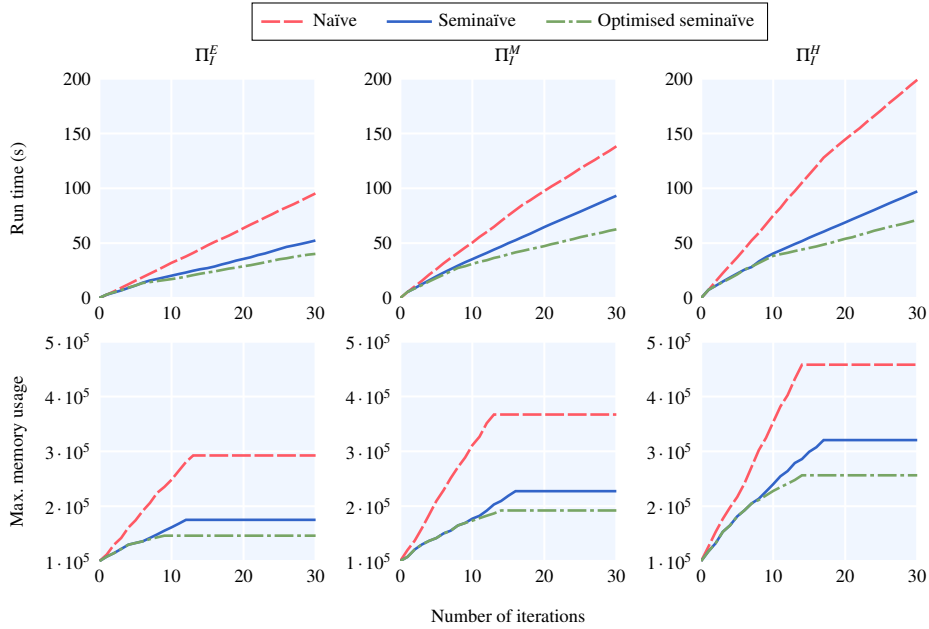

%% file: results/scale_materialisation.tex
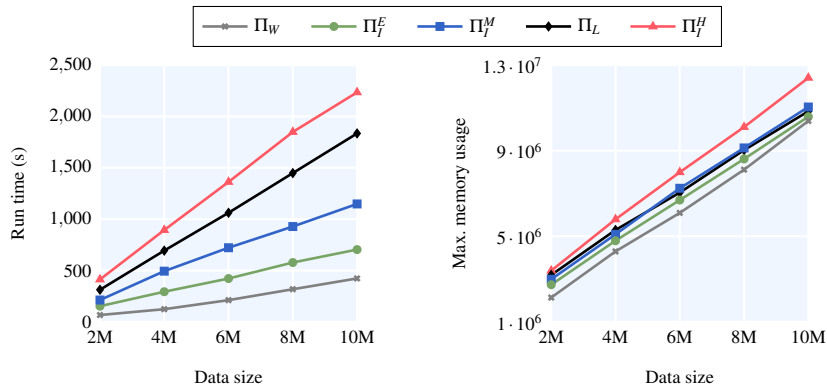
\begin{figure*}[ht]
	\centering
	
\begin{tikzpicture}
\tikzstyle{every node}=[font=\scriptsize]
\begin{axis}[
xshift=0cm,
yshift=0cm,
width = 5cm,
height = 5cm,
xmin=200, xmax=1000,
xtick = {200,400,600,800,1000},
xticklabels = {2M,4M,6M,8M,10M},
ymin = 0, ymax=2500,
ytick={0, 500, 1000, 1500, 2000, 2500},
axis line style={white},
tick style={white},
axis background/.style={fill=Blue4},
grid = both,
major grid style={white,thick},
minor grid style={draw=none},
ylabel={Run time~(s)},
ylabel style={yshift=-0.5em},
xlabel={Data size},
legend style={at={(1.39,1.15)}, anchor=center,align=center},
legend columns=5,
]

\addplot[gray, mark=x, line width=1pt, mark size=1.4pt] table {results/new/mat_scale_weather.txt};

\addplot[Green3,mark=otimes*, line width=1pt, mark size=1.2pt] table
{results/new/mat_scale_itemporal_easy.txt};

\addplot[Blue5, mark=square*, line width=1pt, mark size=1.2pt] table {results/new/mat_scale_itemporal_medium.txt};

\addplot[black, mark=diamond*, line width=1pt, mark size=1.2pt] table  {results/new/mat_scale_lubm.txt};

\addplot[Red1,mark=triangle*, line width=1pt, mark size=1.2pt] table {results/new/mat_scale_itemporal_hard.txt};

\legend{$\Prog_{W}$ \quad , $\Prog_I^{E}$ \quad , $\Prog_I^{M}$ \quad ,  $\Prog_L$ \quad , $\Prog_I^H$
};
\end{axis}

\begin{axis}[
xshift=6cm,
yshift=0cm,
width = 5cm,
height = 5cm,
xmin=200, xmax=1000,
xtick = {200,400,600,800,1000},
xticklabels = {2M,4M,6M,8M,10M},
scaled y ticks=false,
ymin = 1000000, ymax=13000000,
ytick={1000000, 5000000, 9000000, 13000000},
axis line style={white},
tick style={white},
axis background/.style={fill=Blue4},
grid = both,
major grid style={white,thick},
minor grid style={draw=none},
ylabel={Max. memory usage},
ylabel style={yshift=-0.1em},
xlabel={Data size},
]
\addplot[Red1,mark=triangle*, line width=1pt, mark size=1.2pt] table {results/new/mat_scale_itemporal_hard_memory.txt};

\addplot[black, mark=diamond*, line width=1pt, mark size=1.2pt]  table {results/new/mat_scale_lubm_memory.txt};

\addplot[Blue5, mark=square*, line width=1pt, mark size=1.2pt]
 table {results/new/mat_scale_itemporal_medium_memory.txt};

\addplot[Green3,mark=otimes*, line width=1pt, mark size=1.2pt] table {results/new/mat_scale_itemporal_easy_memory.txt};

\addplot[gray, mark=x, line width=1pt, mark size=1.4pt] table {results/new/mat_scale_weather_memory.txt};
\end{axis}
\end{tikzpicture}	
\caption{Scalability of our optimised semina\"ive materialisation}
\label{fig::matscalability}
\end{figure*}

%% file: results/compare_sql.tex
\begin{figure}[ht!]
\centering
\begin{tikzpicture}
\tikzstyle{every node}=[font=\scriptsize]
\begin{axis}[
xshift=0cm,
yshift=0cm,
width = 6cm,
height = 6cm,
xmin=0,xmax=24,
xtick={0,4,...,24},
ymin=0,ymax=24,
ytick={0,4,...,24},
axis line style={white},
tick style={white},
axis background/.style={fill=Blue4},
grid = both,
major grid style={white,thick},
minor grid style={draw=none},
ylabel={Run time (s) for MeTeoR},
ylabel style={yshift=-1.5em},
xlabel={Run time (s) for query rewriting},
xlabel style={yshift=0.5em},
legend style={at={(0.5,1.08)}, anchor=center,align=center},
legend columns=3,
]
\addplot+[scatter, only marks, mark options={fill=none},
scatter/classes={
0={mark=triangle*, Green3, mark size=1.5pt},
1={mark=diamond*, Blue5, mark size=1.5pt},
2={mark=*, Red1, fill=Red1, mark size=1.1pt}
},
scatter src=explicit symbolic] table[x index=0,y index=1,meta index=2,col sep=comma] {results/new/compare_sql_lubm.txt};
\addplot [white, line width=1pt, samples at={0,25}] {x};
\legend{$\Prog_L^{1}$ \quad , $\Prog_L^{2}$ \quad , $\Prog_L^{3}$}
\end{axis}

\begin{axis}[
xshift=7cm,
yshift=0cm,
width = 6cm,
height = 6cm,
xmin=0,xmax=24,
xtick={0,4,...,24},
ymin=0,ymax=24,
ytick={0,4,...,24},
axis line style={white},
tick style={white},
axis background/.style={fill=Blue4},
grid = both,
major grid style={white,thick},
minor grid style={draw=none},
ylabel={Run time (s) for MeTeoR},
ylabel style={yshift=-1.5em},
xlabel={Run time (s) for query rewriting},
xlabel style={yshift=0.5em},
legend style={at={(0.5,1.08)}, anchor=center,align=center},
legend columns=2,
]
\addplot+[scatter, only marks, mark options={fill=none},
scatter/classes={
0={mark=diamond*, Blue5, mark size=1.5pt},
1={mark=*, Red1, fill=Red1, mark size=1.1pt}
},
scatter src=explicit symbolic] table[x index=0,y index=1,meta index=2,col sep=comma] {results/new/compare_sql_weather.txt};
\addplot [white, line width=1pt, samples at={0,25}] {x};
\legend{$\Prog_W^{1}$ \quad , $\Prog_W^{2}$ }
\end{axis}
\end{tikzpicture}		

\caption{Comparison between MeTeoR and the query rewriting baseline}
\label{fig:comparesql}
\end{figure}
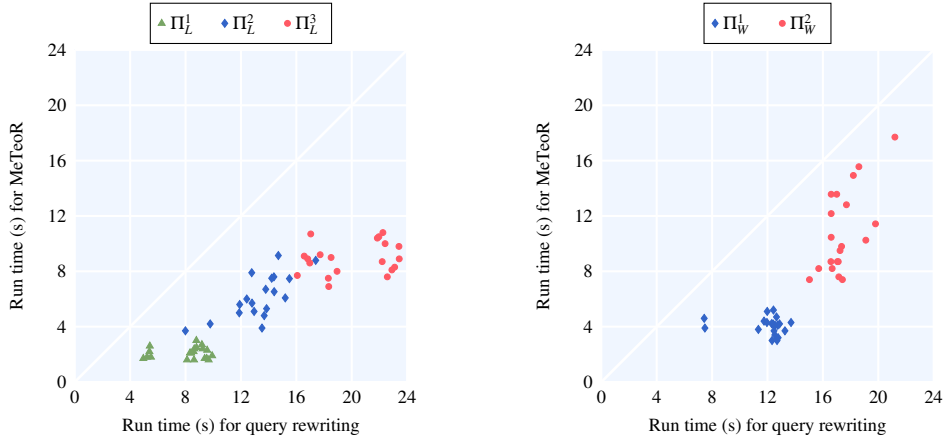

%% file: sections/conclusions.tex
\section{Conclusions}

In this paper, we 
presented a practical reasoning algorithm for full \MTL{}, which combines materialisation with an automata-based approach.
To achieve favourable performance, our algorithm delegates most of the computations to the materialisation component, which we optimised using semina\"ive reasoning strategies.
In turn, the computationally-expensive automata-based approach is used only to guarantee termination and completeness of reasoning.
We have implemented our algorithm in the MeTeoR reasoner and made it  publicly available.
Our extensive evaluation supports our practicality and scalability claims.

We see many  avenues for future research. First, \MTL{} has been 
extended with 
stratified negation-as-failure
\cite{DBLP:conf/aaai/CucalaWGK21} and  our semina\"ive
procedure could be extended accordingly. 
We are also working on blocking conditions that exploit the periodic structure of canonical models to ensure termination of materialisation-based reasoning.
Finally, incremental materialisation-based reasoning has been studied in context of Datalog 
\cite{DBLP:journals/ai/MotikNPH19}, and it would be interesting to lift such approaches to the \MTL{} setting.

%% file: sections/appendix.tex
\section{Proofs}

Theorem~\ref{soundness_n}: Consider Procedure \ref{alg::naive} running on input  $\Prog$ and $\D$. 
Upon the completion of the $k$th (for some $k 
\in \N$) iteration of the loop of  Procedure~\ref{alg::naive}, 
it holds that  $\I_{\D'} \subseteq  T_{\Prog}^{k}(\I_\D)$.
\begin{proof}
For each  $k \in \N$, we let $\mathcal{N}_k$ and $\D_k$ denote the contents of, repsectively, $\mathcal{N}$ and $\D$ in Procedure~\ref{alg::naive}  upon the completion of the $k$th iteration of the loop. Thus, it suffices to show, inductively on $k\in \N$, that $\I_{\D_k} \subseteq T^k_\Prog(\I_\D)$. 

In the base case,  we have $\D_0 = \D$.  Moreover, $T_\Prog^0(\I_\D) = \I_\D$, and so, $\I_{\D_0} \subseteq T_\Prog^0(\I_\D) $, as required. 
For the inductive step, we assume that $\I_{\D_k} \subseteq T_{\Prog}^k(\I_\D)$, for some $k \in \N$, and that the procedure enters the $k+1$st iteration of the loop. If the $k+1$st iteration of the loop breaks in Line~\ref{alg_fixp}, then $\D_{k+1} = \D_{k}$. By the inductive assumption we have $\I_{\D_k} \subseteq T_\Prog^k(\I_\D)$, and therefore, $\I_{\D_{k+1} } \subseteq T_\Prog^k(\I_\D)$, and thus, $\I_{\D_{k+1}} \subseteq T_\Prog^{k+1}(\I_\D)$. 
Now, assume that the $k+1$st iteration of the loop does not break in Line~\ref{alg_fixp}. 
Hence, by Lines~\ref{alg_coal} and \ref{alg_passon}, we have $\D_{k+1} = \coal{\D_{k} \cup \mathcal{N}_{k+1}}$.
To show that  $\I_{\D_{k+1}} \subseteq  T_{\Prog}^{k+1}(\I_\D)$, we assume that  $\I_{\D_{k+1}} \models  M@t$, for some relational fact $M@t$.
Hence, we have  $\D_{k} \models M@t$ or  $\mathcal{N}_{k+1} \models M@t$.
	   
\begin{itemize}
\item[--]  Case 1: \emph{$\D_{k} \models M@t$}. 
By the inductive assumption, 
$\I_{\D_{k}} \subseteq T_{\Prog}^{k}(\I_\D)$,
so $T_{\Prog}^{k}(\I_\D) \models  M@t$.
Clearly,  ${ T_{\Prog}^{k}(\I_\D) \subseteq T_{\Prog}^{k+1}(\I_\D) }$, and so, 
$T_{\Prog}^{k+1}(\I_\D) \models  M@t$, as required.

\item[--] Case 2: \emph{$\mathcal{N}_{k+1} \models M@t$}. 
By Line 3 of Procedure~\ref{alg::naive}, we obtain that $\mathcal{N}_{k+1} = \Prog(\D_{k})$.
Thus, by Expression~\eqref{eq:prog-cons} from Definition~\ref{def::inst}, there exists a rule $r \in \Prog$, say of  the form ${ M' \gets M_1 \land \dots \land M_n }$, such that $r[\D_k] \models M@t$.
Therefore, by  Expression~\eqref{eq:consequences} from Definition~\ref{def::inst}, there are  a substitution $\sigma$ and  some intervals $\varrho_1, \dots, \varrho_n$ such that $(\matA_1\sigma@\varrho_1 , \dots , \matA_n \sigma@\varrho_n) \in \mathsf{inst}_{r}[D_k]$ and
${M'\sigma@(\varrho_1 \cap \dots \cap \varrho_n)\models M@t}$.
Since $(\matA_1\sigma@\varrho_1 , \dots , \matA_n \sigma@\varrho_n)$ belongs to $\mathsf{inst}_{r}[\D_k]$,
by Expression~\eqref{eq:consequences}  from Definition~\ref{def::inst},
we obtain that ${\D_k \models M_i @ \varrho_i}$, for each $i \in \{1, \dots, n \}$. 
Therefore, by the definition of $T_\Prog$, we obtain that ${T_\Prog (\I_{\D_k}) \models M'\sigma@(\varrho_1 \cap \dots \cap \varrho_n) }$ and so $T_\Prog (\I_{\D_k}) \models M@t$.
Finally, by the inductive assumption, ${\I_{\D_k} \subseteq T_\Prog^k(\I_{\D})}$, so
${T_\Prog (\I_{\D_k}) \subseteq T_\Prog^{k+1}(\I_{\D})}$,   	
and therefore we obtain that 
$T_{\Prog}^{k+1}(\I_\D) \models M@t$. 
\end{itemize}
\end{proof}

\noindent Theorem~\ref{complete_n}: Consider Procedure \ref{alg::naive} running on input  $\Prog$ and $\D$. For each $k \in \N$, upon the completion of the $k$th iteration of the loop of  Procedure~\ref{alg::naive},
it holds that  $  T_{\Prog}^{k}(\I_\D) \subseteq \I_{\D'} $.
\begin{proof}
We  use the same symbols $\mathcal{N}_k$ and $\D_k$ as in the proof of Theorem~\ref{soundness_n}
and we define $\Delta_k$ analogously.
We will show, inductively on natural numbers $k$, that $T_\Prog^k(\I_\D) \subseteq \I_{\D_k}$. 
The base case holds trivially, because  we have $\D_0 = \D$,  and so,  $T_\Prog^0(\I_\D) = \I_{\D_0} $.
For the inductive step assume that $T_\Prog^k(\I_\D) \subseteq \I_{\D_k}$, for some $k\in \N$. 
Moreover, let $T_{\Prog}^{k+1}(\I_\D) \models M@t$, for some relational fact $M@t$.
We will show that $\I_{\D_{k+1}} \models  M@t$.
We have either $T_{\Prog}^{k}(\I_\D) \models M@t$ or $T_{\Prog}^{k}(\I_\D) \not\models M@t$.
\begin{itemize}
\item[--]  Case 1:
$T_{\Prog}^{k}(\I_\D) \models M@t$. Then, by the inductive assumption, ${\I_{\D_k} \models M@t}$.
Since $\D_{k+1} \models \D_k$, we obtain that $\I_{\D_{k+1}} \models M@t$.

\item[--]  Case 2: $T_{\Prog}^{k}(\I_\D) \not\models M@t$.
Since 
$T_{\Prog}^{k+1}(\I_\D) \models M@t$,
there exist a rule $r\in \Prog$,
say of the form $M' \gets M_1 \land \dots \land M_n$, and a time point $t'$ such that an application of $r$ at $t'$ yields $M@t$.
More precisely, it means that there is a
substitution $\sigma$
such that $T_{\Prog}^{k}(\I_\D) \models M_i \sigma @t'$, for each ${i \in \{1, \dots, n \} }$, and 
$M'\sigma @ t' \models M@t$.
Therefore,  by the fact that $T_\Prog^k(\I_\D) \subseteq \I_{\D_k} $ and by Expression~\eqref{eq:consequences} from Definition~\ref{def::inst}, there need to exist some intervals $\varrho_1, \dots, \varrho_n$ such that $t' \in \varrho_1 \cap \dots \cap \varrho_n$ and
$(M_1 \sigma @\varrho_1, \cdots, M_n \sigma @\varrho_n) \in \mathsf{inst}_r[\D_k]$.
Thus, as $M'\sigma @ t' \models M@t$, we obtain by Definition~\ref{def::inst} that $ r[\D_k] \models M@t$, and so, $ \Prog[\D_k] \models M@t$.
Finally, recall that $\D_{k+1} = \coal{\D_{k} \cup \mathcal{N}_{k+1}}$ and $\mathcal{N}_{k+1} = \Prog[\D_k]$.
Therefore, the fact that $\Prog[\D_k] \models M@t$ implies that $\I_{\D_{k+1}} \models M@t$.
\end{itemize}
\end{proof}

\noindent Theorem~\ref{soundness_sm}: Consider Procedure \ref{alg::seminaive} running on input  $\Prog$ and $\D$. 
Upon the completion of the $k$th (for some $k 
\in \N$) iteration of the loop of  Procedure~\ref{alg::seminaive}, 
it holds that  $\I_{\D'} \subseteq  T_{\Prog}^{k}(\I_\D)$.
\begin{proof}
The proof relies on the observation that rule instances processed by
semina\"ive evaluation 
are also processed by the na\"ive evaluation.
In particular, directly by  Definition~\ref{def::semi-inst} we obtain that
$\ins{r}{\D' \threedotcolon \Delta} \subseteq \ins{r}{\D'}$, for each $r$, $\D'$, and $\Delta$.
Hence, in each loop iteration, the sets $\mathcal{N}$ and $\mathcal{C}$ in Procedure~\ref{alg::seminaive} are subsets of the corresponding sets in Procedure~\ref{alg::naive}. 
Consequently, the same holds for the set $\D'$, and so, soundness follows from Theorem~\ref{soundness_n}.
\end{proof}

\noindent Theorem~\ref{complete_sm}: Consider Procedure \ref{alg::seminaive} with input program $\Prog$ and input dataset $\D$. For each $k \in \N$, upon the completion of the $k$th iteration of the loop of  Procedure~\ref{alg::seminaive},
it holds that  $  T_{\Prog}^{k}(\I_\D) \subseteq \I_{\D'} $.
\begin{proof}
We will  use  $\mathcal{N}_k$, $\Delta_k$, and  $\D'_k$, for the contents of, respectively, $\mathcal{N}$, $\Delta$, and $\D'$
in Procedure~\ref{alg::seminaive}  upon the completion of the $k$th iteration of the loop. 
Now, as in the proof of Theorem~\ref{complete_n},
we will show inductively on natural numbers $k$, that $T_\Prog^k(\I_\D) \subseteq \I_{\D_k}$.

The only difference with respect to the proof of Theorem~\ref{complete_n} lies in Case 2, where
we  assume that $T_{\Prog}^{k}(\I_\D) \not\models M@t$ and 
$T_{\Prog}^{k+1}(\I_\D) \models M@t$, for some relational fact $M@t$.
By the inductive assumption, as in the proof of Theorem~\ref{complete_n},
there needs to be an instance
$(M_1 \sigma @\varrho_1, \cdots, M_n \sigma @\varrho_n) $ $\in \ins{r}{\D_k}$ of some rule $r$ of the form $M' \gets M_1 \land \dots \land M_n$, and a time point $t' \in \varrho_1 \cap \dots \cap \varrho_n$ such that $M'\sigma @ t' \models M@t$. 
Now, we argue that
this instance is not only in $\ins{r}{\D_k}$, but also in $\ins{r}{\D_k \threedotcolon \Delta_{k} }$, namely that
$(M_1 \sigma @\varrho_1, \cdots, M_n \sigma @\varrho_n) \in \ins{r}{\D_k \threedotcolon \Delta_{k} }$.
To this end, by Definition~\ref{def::semi-inst}, it suffices to show that
there is $i \in \{1, \dots, n \}$ such that ${\D_k \setminus \Delta_k \not\models  M_i \sigma @ \varrho_i}$.
We will consider two cases, namely, when $k=0$ and when $k >0$.
\begin{itemize}
\item[--]  Case 2.1: \emph{$k=0$}. 
By the  initialisation (Line~\ref{alg2_init}) of the of Procedure~\ref{alg::seminaive},$\Delta_0 = \D$, so  $\D_0 \setminus \Delta_0 = \emptyset$.
Recall that we assumed in the paper that input programs do not have rules with vacuously satisfied bodies, so there needs to exist $i \in \{1, \dots, n \}$
such that the empty interpretation does not satisfy
$M_i \sigma @\varrho_i$, that is,
$\D_0 \setminus \Delta_0 \not\models M_i \sigma @\varrho_i$.

\item[--]  Case 2.2: \emph{$k > 0$}. 
By Lines~\ref{alg2_8} and \ref{alg2_coal}, we have $\D_k = \coal{\D_{k-1} \cup \mathcal{N}_k}$.
Moreover, by the definition of $\Delta$ in  Line~\ref{alg2_delta}, we  obtain  $\D_k = \coal{ \D_{k-1} \cup \Delta_k}$.
Thus $\I_{\D_k} = \I_{\D_{k-1}} \cup \I_{\Delta_{k}}$, 
so ${ \I_{\D_k}\setminus \I_{\Delta_k} = \I_{\D_{k-1}} \setminus \I_{\Delta_k} }$, and therefore $\I_{\D_k}\setminus \I_{\Delta_k} \subseteq \I_{\D_{k-1}}$. 
Hence,
to show that for some $i \in \{1, \dots, n \} $ we have that
${\D_k \setminus \Delta_k \not\models  } $ ${M_i \sigma @ \varrho_i}$, it suffices to show that
 $\D_{k-1} \not\models M_i\sigma@\varrho_i$.  
Suppose towards a contradiction that $\D_{k-1} \models M_i\sigma@\varrho_i$, for all $i \in \{1, \dots, n \}$. 
By  Theorem~\ref{soundness_sm},  $T_{\Prog}^{k-1}(\I_{\D}) \models M_i\sigma@\varrho_i$, for all $i \in \{1, \dots, n \}$.
Thus,
$T_{\Prog}^{k}(\I_{\D}) \models M@t$, which raises a contradiction.
\end{itemize}
\end{proof}

\noindent Lemma~\ref{non-rec}:  Consider Procedure~\ref{alg::optseminaive} running on input program $\Prog$ and dataset $\D$.
Whenever  $flag = 1$, dataset $\D'$ satisfies all facts over a 
non-recursive predicate in $\Prog$ that are 
entailed by $\Prog$ and $\D$.
\begin{proof}
In
Procedure~\ref{alg::optseminaive},
$flag$ is initialised to 0 and once it is changed to 1 (in Line \ref{delprognr}) its value cannot be reverted.
Moreover, the
consecutive contents of $\D'$ are obtained by 
coalescing the previous contents with facts in $\mathcal{N}$ (in Line~\ref{line:c}), so once a fact is entailed by
$\D'$, it will remain entailed by all the consecutive contents of $\D'$.
Therefore, to prove the lemma, it suffices to consider the step of computation in which $flag$ becomes 1.
Let $k$ be  this step and let $\D'_k$, be the contents of $\D'$ upon the completion of the $k$th iteration of the loop. 
Hence, we need to show that $\D'_k$ satisfies all the facts
which are satisfied in $\can{\Prog}{\D}$ and which mention non-recursive predicates in $\Prog$. 

For this, we observe that 
before $flag$ changes to 1, Procedure~\ref{alg::optseminaive} works  as Procedure~\ref{alg::seminaive} so,
by Theorems~\ref{soundness_sm} and \ref{complete_sm},
we have ${ \I_{\D'_k} = T_{\Prog}^{k}(\I_\D) }$.
Hence, as $\can{\Prog}{\D} = T_{\Prog}^{\omega_1}(\I_\D)$,  it suffices to  show by a transfinite induction that, 
for each ordinal $\alpha \geq k$, 
the interpretations $T_{\Prog}^{k}(\I_\D)$ and $T_{\Prog}^{\alpha}(\I_\D)$ satisfy the same facts with non-recursive predicates in $\Prog$. 
For the base case assume that $T_{\Prog}^{k+1}(\I_\D) \models M@t$, for some relational fact $M@t$ whose predicate is non-recursive in $\Prog$.
Hence,  there is a rule $r \in \ground{\Prog}{\D}$ and a time point $t'$ such that $T_{\Prog}^{k}(\I_{\D})$ entails 
each body atom of $r$ at $t'$, and the head of $r$ holding at $t'$ entails $M@t$. 
Thus the head atom of $r$ has a non-recursive predicate in $\Prog$; therefore, by the definition, 
each body atom in
$r$ also mentions only 
non-recursive predicates in $\Prog$.
Now, as $T_{\Prog}^{k}(\I_{\D}) \not \models M@t$, we obtain that $T_{\Prog}^{k}(\I_{\D})$ and $T_{\Prog}^{k-1}(\I_{\D})$ do not satisfy the same relational facts  with non-recursive predicates in $\Prog$, which directly contradicts the condition from Line \ref{same}.
The inductive step for a successor ordinal $\alpha$ uses the same argument; indeed, if $T_{\Prog}^{\alpha}(\I_{\D}) \models M@t$ and $T_{\Prog}^{\alpha-1}(\I_{\D}) \models M@t$, for some relational fact $M@t$ whose predicate is non-recursive in $\Prog$, then $T_{\Prog}^{\alpha}(\I_{\D}) $ and $T_{\Prog}^{\alpha-1}(\I_{\D})$ do not satisfy the same relational facts  with non-recursive predicates,  contradicting the inductive assumption.
The inductive step for the transfinite ordinal $\alpha$ holds trivially as $T_{\Prog}^{\alpha} (\I_{\D}) = \bigcup_{\beta < \alpha} T_{\Prog}^{\beta}(\I_{\D}) $.
\end{proof}

\noindent Theorem~\ref{seminaive_theorem}: Consider Procedure \ref{alg::optseminaive} with input program $\Prog$ and input dataset $\D$. For each $k \in \N$, upon the completion of the $k$th iteration of the loop of  Procedure~\ref{alg::seminaive},
it holds that  $  T_{\Prog}^{k}(\I_\D) = \I_{\D'} $.	
\begin{proof}
If for both Lines~\ref{empty} and~\ref{proj}, the condition in the `if' statement never applies, then
Procedure~\ref{alg::optseminaive} works in the same way as Procedure~\ref{alg::seminaive}.
Hence, by Theorems~\ref{soundness_sm} and \ref{complete_sm},
we get  $  T_{\Prog}^{k}(\I_\D) = \I_{\D'} $.
Otherwise, the loop from 
Procedure~\ref{alg::optseminaive} works similarly  as Procedure~\ref{alg::seminaive}, except that it can delete in Lines~\ref{empty}  and \ref{proj} some  rules.
By Lemma~\ref{lemma:invalid}, however, such rules 
can be safely deleted from the program, without losing the properties established in Theorems~\ref{soundness_sm} and \ref{complete_sm}.
\end{proof}

%% file: main.bbl
\begin{thebibliography}{}

\bibitem[\protect\citeauthoryear{Abiteboul, Hull, and Vianu}{Abiteboul
  et~al\mbox{.}}{1995}]{abiteboul1995foundations}
{\sc Abiteboul, S.}, {\sc Hull, R.}, {\sc and} {\sc Vianu, V.} 1995.
\newblock {\em Foundations of Databases}.
\newblock Addison-Wesley.

\bibitem[\protect\citeauthoryear{Aguado, Cabalar, Di\'eguez, P\'erez, Schaub,
  Schuhmann, and Vidal}{Aguado et~al\mbox{.}}{2021}]{aguado2021}
{\sc Aguado, F.}, {\sc Cabalar, P.}, {\sc Di\'eguez, M.}, {\sc P\'erez, G.},
  {\sc Schaub, T.}, {\sc Schuhmann, A.}, {\sc and} {\sc Vidal, C.} 2021.
\newblock Linear-time temporal answer set programming.
\newblock {\em Theory and Practice of Logic Programming\/}, 1–55.

\bibitem[\protect\citeauthoryear{Artale and Franconi}{Artale and
  Franconi}{1998}]{artale1998temporal}
{\sc Artale, A.} {\sc and} {\sc Franconi, E.} 1998.
\newblock A temporal description logic for reasoning about actions and plans.
\newblock {\em Journal of Artificial Intelligence Research\/}, 463--506.

\bibitem[\protect\citeauthoryear{Artale, Kontchakov, Kovtunova, Ryzhikov,
  Wolter, and Zakharyaschev}{Artale et~al\mbox{.}}{2015}]{artale2015first}
{\sc Artale, A.}, {\sc Kontchakov, R.}, {\sc Kovtunova, A.}, {\sc Ryzhikov,
  V.}, {\sc Wolter, F.}, {\sc and} {\sc Zakharyaschev, M.} 2015.
\newblock First-order rewritability of temporal ontology-mediated queries.
\newblock In {\em Proceedings of the International Joint Conference on
  Artificial Intelligence}. 2706--2712.

\bibitem[\protect\citeauthoryear{Baader, Borgwardt, Koopmann, Ozaki, and
  Thost}{Baader et~al\mbox{.}}{2017}]{DBLP:conf/frocos/BaaderBKOT17}
{\sc Baader, F.}, {\sc Borgwardt, S.}, {\sc Koopmann, P.}, {\sc Ozaki, A.},
  {\sc and} {\sc Thost, V.} 2017.
\newblock Metric temporal description logics with interval-rigid names.
\newblock In {\em Frontiers of Combining Systems}. 60--76.

\bibitem[\protect\citeauthoryear{Baier and Katoen}{Baier and
  Katoen}{2008}]{baier2008principles}
{\sc Baier, C.} {\sc and} {\sc Katoen, J.-P.} 2008.
\newblock {\em Principles of model checking}.
\newblock MIT Press.

\bibitem[\protect\citeauthoryear{Beck, Dao-Tran, and Eiter}{Beck
  et~al\mbox{.}}{2018}]{beck2018lars}
{\sc Beck, H.}, {\sc Dao-Tran, M.}, {\sc and} {\sc Eiter, T.} 2018.
\newblock {LARS}: A logic-based framework for analytic reasoning over streams.
\newblock {\em Artificial Intelligence\/}~{\em 261}, 16--70.

\bibitem[\protect\citeauthoryear{Bellomarini, Blasi, Nissl, and
  Sallinger}{Bellomarini et~al\mbox{.}}{2022}]{tempVada}
{\sc Bellomarini, L.}, {\sc Blasi, L.}, {\sc Nissl, M.}, {\sc and} {\sc
  Sallinger, E.} 2022.
\newblock The temporal {V}adalog system.
\newblock In {\em International Joint Conference on Rules and Reasoning}.
  130--145.

\bibitem[\protect\citeauthoryear{Bellomarini, Nissl, and Sallinger}{Bellomarini
  et~al\mbox{.}}{2021}]{bellomarini2021query}
{\sc Bellomarini, L.}, {\sc Nissl, M.}, {\sc and} {\sc Sallinger, E.} 2021.
\newblock Query evaluation in {DatalogMTL} -- taming infinite query results.
\newblock {\em CoRR\/}~{\em abs/2109.10691}.

\bibitem[\protect\citeauthoryear{Bellomarini, Nissl, and Sallinger}{Bellomarini
  et~al\mbox{.}}{2022}]{iTemporal}
{\sc Bellomarini, L.}, {\sc Nissl, M.}, {\sc and} {\sc Sallinger, E.} 2022.
\newblock {iTemporal}: {A}n extensible temporal benchmark generator.
\newblock In {\em 2022 IEEE 38th International Conference on Data Engineering}.
  2021--2033.

\bibitem[\protect\citeauthoryear{Bellomarini, Sallinger, and
  Gottlob}{Bellomarini
  et~al\mbox{.}}{2018}]{DBLP:journals/pvldb/BellomariniSG18}
{\sc Bellomarini, L.}, {\sc Sallinger, E.}, {\sc and} {\sc Gottlob, G.} 2018.
\newblock The {V}adalog system: Datalog-based reasoning for knowledge graphs.
\newblock {\em Proceedings of the VLDB Endowment\/}, 975--987.

\bibitem[\protect\citeauthoryear{Brandt, Kalayc{\i}, Kontchakov, Ryzhikov,
  Xiao, and Zakharyaschev}{Brandt et~al\mbox{.}}{2017}]{brandt2017ontology}
{\sc Brandt, S.}, {\sc Kalayc{\i}, E.~G.}, {\sc Kontchakov, R.}, {\sc Ryzhikov,
  V.}, {\sc Xiao, G.}, {\sc and} {\sc Zakharyaschev, M.} 2017.
\newblock Ontology-based data access with a horn fragment of metric temporal
  logic.
\newblock In {\em Proceedings of the {AAAI} Conference on Artificial
  Intelligence}. 1070--1076.

\bibitem[\protect\citeauthoryear{Brandt, Kalayc{\i}, Ryzhikov, Xiao, and
  Zakharyaschev}{Brandt et~al\mbox{.}}{2018}]{brandt2018querying}
{\sc Brandt, S.}, {\sc Kalayc{\i}, E.~G.}, {\sc Ryzhikov, V.}, {\sc Xiao, G.},
  {\sc and} {\sc Zakharyaschev, M.} 2018.
\newblock Querying log data with metric temporal logic.
\newblock {\em Journal of Artificial Intelligence Research\/}~{\em 62},
  829--877.

\bibitem[\protect\citeauthoryear{Bry, Eisinger, Eiter, Furche, Gottlob, Ley,
  Linse, Pichler, and Wei}{Bry
  et~al\mbox{.}}{2007}]{DBLP:conf/rweb/BryEEFGLLPW07}
{\sc Bry, F.}, {\sc Eisinger, N.}, {\sc Eiter, T.}, {\sc Furche, T.}, {\sc
  Gottlob, G.}, {\sc Ley, C.}, {\sc Linse, B.}, {\sc Pichler, R.}, {\sc and}
  {\sc Wei, F.} 2007.
\newblock Foundations of rule-based query answering.
\newblock In {\em Reasoning Web}. 1--153.

\bibitem[\protect\citeauthoryear{Cabalar, Di\'eguez, Schaub, and
  Schuhmann}{Cabalar
  et~al\mbox{.}}{2020}]{cabalar_dieguez_schaub_schuhmann_2020}
{\sc Cabalar, P.}, {\sc Di\'eguez, M.}, {\sc Schaub, T.}, {\sc and} {\sc
  Schuhmann, A.} 2020.
\newblock Towards metric temporal answer set programming.
\newblock {\em Theory and Practice of Logic Programming\/}~{\em 20,\/}~5,
  783--798.

\bibitem[\protect\citeauthoryear{Carral, Dragoste, Gonz{\'{a}}lez, Jacobs,
  Kr{\"{o}}tzsch, and Urbani}{Carral
  et~al\mbox{.}}{2019}]{DBLP:conf/semweb/CarralDGJKU19}
{\sc Carral, D.}, {\sc Dragoste, I.}, {\sc Gonz{\'{a}}lez, L.}, {\sc Jacobs, C.
  J.~H.}, {\sc Kr{\"{o}}tzsch, M.}, {\sc and} {\sc Urbani, J.} 2019.
\newblock Vlog: a rule engine for knowledge graphs.
\newblock In {\em International Semantic Web Conference}. 19--35.

\bibitem[\protect\citeauthoryear{Cavada, Cimatti, Dorigatti, Griggio, Mariotti,
  Micheli, Mover, Roveri, and Tonetta}{Cavada
  et~al\mbox{.}}{2014}]{DBLP:conf/cav/CavadaCDGMMMRT14}
{\sc Cavada, R.}, {\sc Cimatti, A.}, {\sc Dorigatti, M.}, {\sc Griggio, A.},
  {\sc Mariotti, A.}, {\sc Micheli, A.}, {\sc Mover, S.}, {\sc Roveri, M.},
  {\sc and} {\sc Tonetta, S.} 2014.
\newblock The nuxmv symbolic model checker.
\newblock In {\em Computer Aided Verification}. 334--342.

\bibitem[\protect\citeauthoryear{Ceri, Gottlob, and Tanca}{Ceri
  et~al\mbox{.}}{1989}]{ceri1989you}
{\sc Ceri, S.}, {\sc Gottlob, G.}, {\sc and} {\sc Tanca, L.} 1989.
\newblock What you always wanted to know about {D}atalog (and never dared to
  ask).
\newblock {\em IEEE Transactions on Knowledge and Data Engineering\/}~{\em
  1,\/}~1, 146--166.

\bibitem[\protect\citeauthoryear{Chomicki and Imielinski}{Chomicki and
  Imielinski}{1988}]{chomicki1988temporal}
{\sc Chomicki, J.} {\sc and} {\sc Imielinski, T.} 1988.
\newblock Temporal deductive databases and infinite objects.
\newblock In {\em Proceedings of the Seventh {ACM} {SIGACT-SIGMOD-SIGART}
  Symposium on Principles of Database Systems}. 61--73.

\bibitem[\protect\citeauthoryear{Cucala, Wa{\l}{\k{e}}ga, {Cuenca Grau}, and
  Kostylev}{Cucala et~al\mbox{.}}{2021}]{DBLP:conf/aaai/CucalaWGK21}
{\sc Cucala, D. J.~T.}, {\sc Wa{\l}{\k{e}}ga, P.~A.}, {\sc {Cuenca Grau}, B.},
  {\sc and} {\sc Kostylev, E.~V.} 2021.
\newblock Stratified negation in {D}atalog with metric temporal operators.
\newblock In {\em Proceedings of the Thirty-Third {AAAI} Conference on
  Artificial Intelligence}. 6488--6495.

\bibitem[\protect\citeauthoryear{Eiter, Leone, Mateis, Pfeifer, and
  Scarcello}{Eiter et~al\mbox{.}}{1998}]{DBLP:conf/kr/EiterLMPS98}
{\sc Eiter, T.}, {\sc Leone, N.}, {\sc Mateis, C.}, {\sc Pfeifer, G.}, {\sc
  and} {\sc Scarcello, F.} 1998.
\newblock The {KR} system dlv: Progress report, comparisons and benchmarks.
\newblock In {\em Proceedings of the Sixth International Conference on
  Principles of Knowledge Representation and Reasoning (KR'98), Trento, Italy,
  June 2-5, 1998}, {A.~G. Cohn}, {L.~K. Schubert}, {and} {S.~C. Shapiro}, Eds.
  Morgan Kaufmann, 406--417.

\bibitem[\protect\citeauthoryear{Geatti, Gigante, and Montanari}{Geatti
  et~al\mbox{.}}{2019}]{DBLP:conf/tableaux/GeattiGM19}
{\sc Geatti, L.}, {\sc Gigante, N.}, {\sc and} {\sc Montanari, A.} 2019.
\newblock A sat-based encoding of the one-pass and tree-shaped tableau system
  for {LTL}.
\newblock In {\em Automated Reasoning with Analytic Tableaux and Related
  Methods}. 3--20.

\bibitem[\protect\citeauthoryear{Gebser, Kaminski, Kaufmann, and Schaub}{Gebser
  et~al\mbox{.}}{2014}]{DBLP:journals/corr/GebserKKS14}
{\sc Gebser, M.}, {\sc Kaminski, R.}, {\sc Kaufmann, B.}, {\sc and} {\sc
  Schaub, T.} 2014.
\newblock Clingo = {ASP} + control: Preliminary report.
\newblock {\em CoRR\/}~{\em abs/1405.3694}.

\bibitem[\protect\citeauthoryear{Gebser, Kaufmann, and Schaub}{Gebser
  et~al\mbox{.}}{2012}]{DBLP:journals/ai/GebserKS12}
{\sc Gebser, M.}, {\sc Kaufmann, B.}, {\sc and} {\sc Schaub, T.} 2012.
\newblock Conflict-driven answer set solving: From theory to practice.
\newblock {\em Artif. Intell.\/}~{\em 187}, 52--89.

\bibitem[\protect\citeauthoryear{Guo, Pan, and Heflin}{Guo
  et~al\mbox{.}}{2005}]{guo2005lubm}
{\sc Guo, Y.}, {\sc Pan, Z.}, {\sc and} {\sc Heflin, J.} 2005.
\newblock {LUBM}: a benchmark for {OWL} knowledge base systems.
\newblock {\em Journal of Web Semantics\/}~{\em 3,\/}~2-3, 158--182.

\bibitem[\protect\citeauthoryear{Guti{\'{e}}rrez{-}Basulto, Jung, and
  Ozaki}{Guti{\'{e}}rrez{-}Basulto et~al\mbox{.}}{2016}]{gutierrez2016metric}
{\sc Guti{\'{e}}rrez{-}Basulto, V.}, {\sc Jung, J.~C.}, {\sc and} {\sc Ozaki,
  A.} 2016.
\newblock On metric temporal description logics.
\newblock In {\em 22nd European Conference on Artificial Intelligence}.
  837--845.

\bibitem[\protect\citeauthoryear{Kalayc{\i}, Xiao, Ryzhikov, Kalayci, and
  Calvanese}{Kalayc{\i} et~al\mbox{.}}{2018}]{guzel2018ontop}
{\sc Kalayc{\i}, E.~G.}, {\sc Xiao, G.}, {\sc Ryzhikov, V.}, {\sc Kalayci,
  T.~E.}, {\sc and} {\sc Calvanese, D.} 2018.
\newblock Ontop-temporal: a tool for ontology-based query answering over
  temporal data.
\newblock In {\em Proceedings of the 27th ACM International Conference on
  Information and Knowledge Management}. 1927--1930.

\bibitem[\protect\citeauthoryear{Kikot, Ryzhikov, Wa{\l}{\k{e}}ga, and
  Zakharyaschev}{Kikot et~al\mbox{.}}{2018}]{kikot2018data}
{\sc Kikot, S.}, {\sc Ryzhikov, V.}, {\sc Wa{\l}{\k{e}}ga, P.~A.}, {\sc and}
  {\sc Zakharyaschev, M.} 2018.
\newblock On the data complexity of ontology-mediated queries with {MTL}
  operators over timed words.
\newblock In {\em Description Logics}.

\bibitem[\protect\citeauthoryear{Kontchakov, Pandolfo, Pulina, Ryzhikov, and
  Zakharyaschev}{Kontchakov et~al\mbox{.}}{2016}]{10.5555/3060621.3060782}
{\sc Kontchakov, R.}, {\sc Pandolfo, L.}, {\sc Pulina, L.}, {\sc Ryzhikov, V.},
  {\sc and} {\sc Zakharyaschev, M.} 2016.
\newblock Temporal and spatial {OBDA} with many-dimensional {H}alpern-{S}hoham
  logic.
\newblock In {\em Proceedings of the International Joint Conference on
  Artificial Intelligence}. 1160--1166.

\bibitem[\protect\citeauthoryear{Koopmann}{Koopmann}{2019}]{koopmann2019ontology}
{\sc Koopmann, P.} 2019.
\newblock Ontology-based query answering for probabilistic temporal data.
\newblock In {\em Proceedings of the AAAI Conference on Artificial
  Intelligence}. 2903--2910.

\bibitem[\protect\citeauthoryear{Koymans}{Koymans}{1990}]{koymans1990specifying}
{\sc Koymans, R.} 1990.
\newblock Specifying real-time properties with metric temporal logic.
\newblock {\em Real-time Systems\/}~{\em 2,\/}~4, 255--299.

\bibitem[\protect\citeauthoryear{Maurer, Wood, Adam, Lettenmaier, and
  Nijssen}{Maurer et~al\mbox{.}}{2002}]{maurer2002long}
{\sc Maurer, E.~P.}, {\sc Wood, A.~W.}, {\sc Adam, J.~C.}, {\sc Lettenmaier,
  D.~P.}, {\sc and} {\sc Nijssen, B.} 2002.
\newblock A long-term hydrologically based dataset of land surface fluxes and
  states for the conterminous united states.
\newblock {\em Journal of Climate\/}~{\em 15,\/}~22, 3237--3251.

\bibitem[\protect\citeauthoryear{Mori, Papotti, Bellomarini, and Giudice}{Mori
  et~al\mbox{.}}{2022}]{mori2022neural}
{\sc Mori, M.}, {\sc Papotti, P.}, {\sc Bellomarini, L.}, {\sc and} {\sc
  Giudice, O.} 2022.
\newblock Neural machine translation for fact-checking temporal claims.
\newblock In {\em Proceedings of the Fifth Fact Extraction and VERification
  Workshop}. 78--82.

\bibitem[\protect\citeauthoryear{Motik, Nenov, Piro, and Horrocks}{Motik
  et~al\mbox{.}}{2019}]{DBLP:journals/ai/MotikNPH19}
{\sc Motik, B.}, {\sc Nenov, Y.}, {\sc Piro, R.}, {\sc and} {\sc Horrocks, I.}
  2019.
\newblock Maintenance of {D}atalog materialisations revisited.
\newblock {\em Artificial Intelligence\/}~{\em 269}, 76--136.

\bibitem[\protect\citeauthoryear{Motik, Nenov, Piro, Horrocks, and
  Olteanu}{Motik et~al\mbox{.}}{2014}]{motik2014parallel}
{\sc Motik, B.}, {\sc Nenov, Y.}, {\sc Piro, R.}, {\sc Horrocks, I.}, {\sc and}
  {\sc Olteanu, D.} 2014.
\newblock Parallel materialisation of {D}atalog programs in centralised,
  main-memory {RDF} systems.
\newblock In {\em Proceedings of the AAAI Conference on Artificial
  Intelligence}.

\bibitem[\protect\citeauthoryear{Nissl and Sallinger}{Nissl and
  Sallinger}{2022}]{DBLP:conf/edbt/NisslS22}
{\sc Nissl, M.} {\sc and} {\sc Sallinger, E.} 2022.
\newblock Modelling smart contracts with datalogmtl.
\newblock In {\em Proceedings of the Workshops of the {EDBT/ICDT}}.

\bibitem[\protect\citeauthoryear{Raheb, Mailis, Ryzhikov, Papapetrou, and
  Ioannidis}{Raheb et~al\mbox{.}}{2017}]{raheb2017balonse}
{\sc Raheb, K.~E.}, {\sc Mailis, T.}, {\sc Ryzhikov, V.}, {\sc Papapetrou, N.},
  {\sc and} {\sc Ioannidis, Y.~E.} 2017.
\newblock Balonse: Temporal aspects of dance movement and its ontological
  representation.
\newblock In {\em European Semantic Web Conference}. 49--64.

\bibitem[\protect\citeauthoryear{Ryzhikov, Wa{\l}{\k{e}}ga, and
  Zakharyaschev}{Ryzhikov et~al\mbox{.}}{2019}]{ryzhikov2019data}
{\sc Ryzhikov, V.}, {\sc Wa{\l}{\k{e}}ga, P.~A.}, {\sc and} {\sc Zakharyaschev,
  M.} 2019.
\newblock Data complexity and rewritability of ontology-mediated queries in
  metric temporal logic under the event-based semantics.
\newblock In {\em International Joint Conferences on Artificial Intelligence}.
  1851--1857.

\bibitem[\protect\citeauthoryear{Thost}{Thost}{2018}]{thost2018metric}
{\sc Thost, V.} 2018.
\newblock Metric temporal extensions of {DL-Lite} and interval-rigid names.
\newblock In {\em Proceedings of the International Conference on Principles of
  Knowledge Representation and Reasoning}. 665--666.

\bibitem[\protect\citeauthoryear{Wa{\l}{\k{e}}ga, Zawidzki, and
  Grau}{Wa{\l}{\k{e}}ga et~al\mbox{.}}{2023}]{walkega2023finite}
{\sc Wa{\l}{\k{e}}ga, P.}, {\sc Zawidzki, M.}, {\sc and} {\sc Grau, B.~C.}
  2023.
\newblock Finite materialisability of {D}atalog programs with metric temporal
  operators.
\newblock {\em Journal of Artificial Intelligence Research\/}~{\em 76},
  471--521.

\bibitem[\protect\citeauthoryear{Wa{\l}{\k{e}}ga, {Cuenca Grau}, Kaminski, and
  Kostylev}{Wa{\l}{\k{e}}ga et~al\mbox{.}}{2019}]{walega2019datalogmtl}
{\sc Wa{\l}{\k{e}}ga, P.~A.}, {\sc {Cuenca Grau}, B.}, {\sc Kaminski, M.}, {\sc
  and} {\sc Kostylev, E.~V.} 2019.
\newblock {DatalogMTL}: computational complexity and expressive power.
\newblock In {\em International Joint Conferences on Artificial Intelligence}.
  1886--1892.

\bibitem[\protect\citeauthoryear{Wa{\l}{\k{e}}ga, {Cuenca Grau}, Kaminski, and
  Kostylev}{Wa{\l}{\k{e}}ga et~al\mbox{.}}{2020}]{DBLP:conf/kr/WalegaGKK20}
{\sc Wa{\l}{\k{e}}ga, P.~A.}, {\sc {Cuenca Grau}, B.}, {\sc Kaminski, M.}, {\sc
  and} {\sc Kostylev, E.~V.} 2020.
\newblock Datalog{MTL} over the integer timeline.
\newblock In {\em Proceedings of the International Conference on Principles of
  Knowledge Representation and Reasoning}. 768--777.

\bibitem[\protect\citeauthoryear{Wa{\l}{\k{e}}ga, Cuenca~Grau, Kaminski, and
  Kostylev}{Wa{\l}{\k{e}}ga et~al\mbox{.}}{2021}]{DBLP:conf/ijcai/WalegaGKK20}
{\sc Wa{\l}{\k{e}}ga, P.~A.}, {\sc Cuenca~Grau, B.}, {\sc Kaminski, M.}, {\sc
  and} {\sc Kostylev, E.~V.} 2021.
\newblock Tractable fragments of {D}atalog with metric temporal operators.
\newblock In {\em Proceedings of the AAAI Conference on Artificial
  Intelligence}. 1919--1925.

\bibitem[\protect\citeauthoryear{Wa{\l}{\k{e}}ga, Kaminski, and
  Cuenca~Grau}{Wa{\l}{\k{e}}ga et~al\mbox{.}}{2019}]{walega2019reasoning}
{\sc Wa{\l}{\k{e}}ga, P.~A.}, {\sc Kaminski, M.}, {\sc and} {\sc Cuenca~Grau,
  B.} 2019.
\newblock Reasoning over streaming data in metric temporal {D}atalog.
\newblock In {\em Proceedings of the AAAI Conference on Artificial
  Intelligence}. 3092--3099.

\bibitem[\protect\citeauthoryear{Wa{\l}{\k{e}}ga, Zawidzki, and
  Cuenca~Grau}{Wa{\l}{\k{e}}ga et~al\mbox{.}}{2021}]{walega2021finitely}
{\sc Wa{\l}{\k{e}}ga, P.~A.}, {\sc Zawidzki, M.}, {\sc and} {\sc Cuenca~Grau,
  B.} 2021.
\newblock Finitely materialisable {D}atalog programs with metric temporal
  operators.
\newblock In {\em Proceedings of the International Conference on Principles of
  Knowledge Representation and Reasoning}. 619--628.

\bibitem[\protect\citeauthoryear{Wa\l\k{e}ga, {Tena Cucala}, Kostylev, and
  {Cuenca Grau}}{Wa\l\k{e}ga et~al\mbox{.}}{2021}]{walkega2021datalogmtl}
{\sc Wa\l\k{e}ga, P.~A.}, {\sc {Tena Cucala}, D.~J.}, {\sc Kostylev, E.~V.},
  {\sc and} {\sc {Cuenca Grau}, B.} 2021.
\newblock {DatalogMTL} with negation under stable models semantics.
\newblock In {\em Proceedings of the International Conference on Principles of
  Knowledge Representation and Reasoning}. 609--618.

\bibitem[\protect\citeauthoryear{Wang, Hu, Walega, and {Cuenca Grau}}{Wang
  et~al\mbox{.}}{2022}]{wang2022meteor}
{\sc Wang, D.}, {\sc Hu, P.}, {\sc Walega, P.~A.}, {\sc and} {\sc {Cuenca
  Grau}, B.} 2022.
\newblock Meteor: Practical reasoning in datalog with metric temporal
  operators.
\newblock In {\em Thirty-Sixth {AAAI} Conference on Artificial Intelligence,
  ({AAAI}-2022)}. {AAAI} Press, 5906--5913.

\bibitem[\protect\citeauthoryear{Wang, Walega, and {Cuenca Grau}}{Wang
  et~al\mbox{.}}{2022}]{seminaiveRR}
{\sc Wang, D.}, {\sc Walega, P.~A.}, {\sc and} {\sc {Cuenca Grau}, B.} 2022.
\newblock Semina{\"{\i}}ve materialisation in datalogmtl.
\newblock In {\em Rules and Reasoning - 6th International Joint Conference on
  Rules and Reasoning, RuleML+RR 2022, Berlin, Germany, September 26-28, 2022,
  Proceedings}, {G.~Governatori} {and} {A.~Turhan}, Eds. Lecture Notes in
  Computer Science, vol. 13752. Springer, 183--197.

\bibitem[\protect\citeauthoryear{Yang}{Yang}{2022}]{Yang-2022}
{\sc Yang, J.} 2022.
\newblock Translation of {DatalogMTL} into {PLTL}.
\newblock M.S.\ thesis, University of Oxford, UK.

\bibitem[\protect\citeauthoryear{Zhou, Wang, and Zaniolo}{Zhou
  et~al\mbox{.}}{2006}]{zhou2006efficient}
{\sc Zhou, X.}, {\sc Wang, F.}, {\sc and} {\sc Zaniolo, C.} 2006.
\newblock Efficient temporal coalescing query support in relational database
  systems.
\newblock In {\em International Conference on Database and Expert Systems
  Applications}. 676--686.

\end{thebibliography}
